\documentclass[acmsmall]{acmart}
\usepackage{booktabs} % For formal tables
\usepackage{enumitem}
\usepackage{hyperref}
\usepackage{graphicx}
\usepackage{amsmath}
\usepackage{amsfonts}
\usepackage{subfigure}
\usepackage{algorithmic}
\usepackage{multirow}
\usepackage{xcolor}
\usepackage[ruled]{algorithm2e} % For algorithms

\newcommand{\ie}{\emph{i.e., }}
\newcommand{\eg}{\emph{e.g., }}

\newcommand{\dxy}[1]{#1}
\AtBeginDocument{%
  \providecommand\BibTeX{{%
    \normalfont B\kern-0.5em{\scshape i\kern-0.25em b}\kern-0.8em\TeX}}}

\setcopyright{acmcopyright}
\acmJournal{TOIS}
\acmYear{2019}\acmVolume{0}\acmNumber{0}\acmArticle{111}\acmMonth{0}
\acmDOI{10.1145/XXXXXXXXXX}

\begin{document}

%
% The "title" command has an optional parameter, allowing the author to define a "short title" to be used in page headers.
\title[Convolutional NCF]{Modeling Embedding Dimension Correlations via Convolutional Neural Collaborative Filtering}

%
% The "author" command and its associated commands are used to define the authors and their affiliations.
% Of note is the shared affiliation of the first two authors, and the "authornote" and "authornotemark" commands
% used to denote shared contribution to the research.
\author{Xiaoyu~Du}
\email{duxy.me@gmail.com}
\affiliation{%
  \institution{University of Electronic Science and Technology of China}
  \city{Chengdu}
  \state{Sichuan}
  \country{China}
}

\author{Xiangnan~He}
\authornote{Xiangnan He is the corresponding author.}
\email{xiangnanhe@gmail.com}
\affiliation{%
  \institution{University of Science and Technology of China}
  \city{Hefei}
  \state{Anhui}
  \country{China}
}

\author{Fajie~Yuan}
\email{fajieyuan@tencent.com}
\affiliation{%
  \institution{Platform and Content Group (PCG) of Tencent}
  \city{Shenzhen}
  \state{Guangdong}
  \country{China}
}

\author{Jinhui~Tang}
\email{jinhuitang@njust.edu.cn}
\affiliation{%
  \institution{Nanjing University of Science and Technology}
  \city{Nanjing}
  \state{Jiangsu}
  \country{China}
}

\author{Zhiguang~Qin}
\email{qinzg@uestc.edu.cn}
\affiliation{%
  \institution{University of Electronic Science and Technology of China}
  \city{Chengdu}
  \state{Sichuan}
  \country{China}
}
\author{Tat-Seng~Chua}
\email{dcscts@nus.edu.sg}
\affiliation{%
  \institution{National University of Singapore}
  \city{Singapore}
  \country{Singapore}
}
%
% By default, the full list of authors will be used in the page headers. Often, this list is too long, and will overlap
% other information printed in the page headers. This command allows the author to define a more concise list
% of authors' names for this purpose.
\renewcommand{\shortauthors}{Xiaoyu Du, et al.}

%
% The abstract is a short summary of the work to be presented in the article.
\begin{abstract}
	As the core of recommender system, collaborative filtering (CF) models the affinity between a user and an item from historical user-item interactions, such as clicks, purchases, and so on. Benefited from the strong representation power, neural networks have recently revolutionized the recommendation research, setting up a new standard for CF. However, existing neural recommender models do not explicitly consider the correlations among embedding dimensions, making them less effective in modeling the interaction function between users and items. In this work, we emphasize on modeling the correlations among embedding dimensions in neural networks to pursue higher effectiveness for CF. We propose a novel and general neural collaborative filtering framework, namely ConvNCF, which is featured with two designs: 1) applying outer product on user embedding and item embedding to explicitly model the pairwise correlations between embedding dimensions, and 2) employing convolutional neural network above the outer product to learn the high-order correlations among embedding dimensions. To justify our proposal, we present three instantiations of ConvNCF by using different inputs to represent a user and conduct experiments on two real-world datasets. Extensive results verify the utility of modeling embedding dimension correlations with ConvNCF, which outperforms several competitive CF methods.

\end{abstract}



%
% Keywords. The author(s) should pick words that accurately describe the work being
% presented. Separate the keywords with commas.

%
% A "teaser" image appears between the author and affiliation information and the body 
% of the document, and typically spans the page. 
%%  \includegraphics[width=\textwidth]{sampleteaser}
%%  \caption{Seattle Mariners at Spring Training, 2010.}
%%  \Description{Enjoying the baseball game from the third-base seats. Ichiro Suzuki preparing to bat.}

%
% This command processes the author and affiliation and title information and builds
% the first part of the formatted document.
\maketitle

\section{Introduction}
\label{sec:intro}
% Background 
Recommender system plays a vital role in the era of information explosion. It not only alleviates the information overload issue and facilities the information seeking for users, but also serves as an effective solution to increase the traffic and revenue for service providers. Given its extensive use in Web applications like E-commerce~\cite{Yu:2018:ACR}, social media~\cite{wang2017item}, news portal~\cite{DKN}, and music sites~\cite{cao2017embedding}, its importance cannot be overstated as a highly valuable learning system. For example, it is reported that about $30\%$ traffic in Amazon are brought by recommendations~\cite{sharma2015estimating}, and the Netflix recommender system has contributed over \$1 billion revenue per year~\cite{gomez2016netflix}.

% CF Model, inner product
Modern recommender systems typically rely on collaborative filtering (and/or its content/context -aware variants) to predict a user's preference on unknown items~\cite{iCD,he2019fast,deng2017deep,zhao2018automatically}. The learning objective can be abstracted as estimating the affinity score between a user and an item, such that the recommendation list for a user can be obtained by ranking the candidate items according to the affinity scores. Generally speaking, the key to build a CF model lies in twofold: 1) how to represent a user and an item, and 2) how to build the predictive function based on user and item representations. As a pioneering CF model, matrix factorization (MF)~\cite{koren2008factorization} embeds a user and an item into vector representations, and formulates the predictive function as the inner product between user embedding vector item embedding vector.
%\begin{equation}
%    f_{MF}(u,i) = \textbf{p}_u \cdot \textbf{q}_i = \sum_{k=1}^K p_{uk}q_{ik}, 
%\end{equation}
%where $\textbf{p}_u \in \mathbb{R}^K$ and $\textbf{q}_i \in \mathbb{R}^K$ denote the embedding vector for user $u$ and item $i$, respectively. 
Soon after MF was introduced to recommender system, it becomes prevalent in recommendation research with many variants developed. Some representative  variants include FISM~(factored item similarity model)~\cite{kabbur2013fism} and SVD++~\cite{koren2008factorization} which enrich the user embedding with the embeddings of the user's interacted items, and FM~(factorization machine)~\cite{rendle2010factorization} and SVDfeature~\cite{chen2012svdfeature,yuan2018fbgd} which extend the predictive function with the inner product between the embeddings of content (and/or context) features. 

% Move to neural
Without exaggeration, we would say that factorization methods are the most popular approach and play a dominant role in recommendation research in the past decade. In essence, MF leads to the \textit{de facto} standard for modeling the CF effect --- measuring the user-item affinity with inner product in the embedding space. While inner product is effective in capturing the low-rank structure in sparse user-item interaction data~\cite{LatentCross}, its simplicity (i.e., no learnable parameters are involved) and linearity (i.e., no any nonlinear transformations) limit the expressive power of the predictive function. For example, He et al.~\cite{he2017neural} demonstrated that it may lead to unexpected ranking loss when the embedding size is not sufficiently large, and Hsieh et al.~\cite{hsieh2017collaborative} illustrated the cases that inner product may fail due to its violation of the triangle inequality. 

\begin{figure*}[t]
	\includegraphics[width=\linewidth]{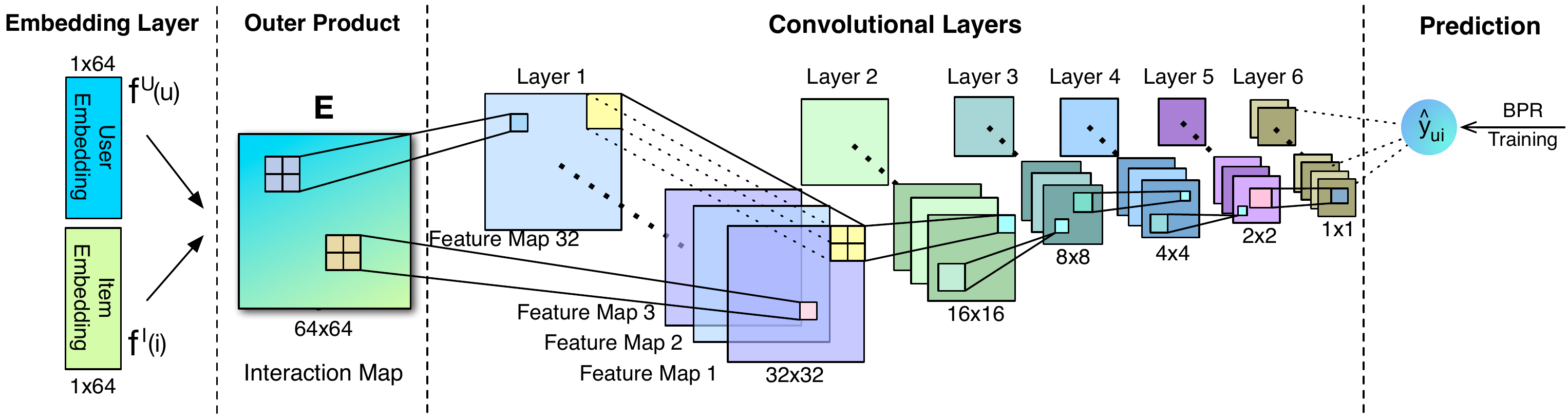}
	\caption{An illustration of our proposed \textbf{Conv}olutional \textbf{N}eural \textbf{C}ollaborative \textbf{F}iltering (ConvNCF) solution. Following the embedding layer is an outer product layer, which generates a 2D matrix (interaction map) that explicitly captures the pairwise correlations between embedding dimensions. The interaction map is then fed into a CNN to model high-order correlations to obtain the final prediction.}
	\label{fig:newframe}
\end{figure*}

% Argue limitations of existing neural RS models
To address the limitations of factorization methods and move towards the next generation of recommender systems, neural network models have been explored for CF in recent years. To date, most neural recommender models can be described within the neural collaborative filtering (NCF) framework~\cite{he2017neural}, which use neural networks to emphasize on either the user/item representation learning part~\cite{DeepMF,wu2016collaborative,he2016vbpr} or the predictive function part~\cite{zhang2017joint,ACF,yuan2019simple}. Although these methods have achieved substantial improvements, we argue a common drawback is that, they forgo considering the correlations among embedding dimensions in the predictive function. To be specific, the mainstream design of the predictive function is to place a multi-layer perceptron (MLP) above the concatenation~(which retains the original information)~\cite{he2017neural} or element-wise product~(which subsumes the inner product)~\cite{zhang2017joint} of user embedding and item embedding. Apparently, both operations assume the embedding dimensions are independent of each other, that is, no correlations among the dimensions are captured. Arguably, the following MLP network can approximate any continuous function~\cite{cybenko1989approximation}, therefore it might be capable of capturing the possible correlations. However, such process is rather implicit and as such, it may be ineffective in capturing certain relations --- an evidence is from \cite{LatentCross} showing that much more parameters have to be used in order to approximate the simple multiplicative relation.

In this work, we highlight the importance of modeling the correlations among embedding dimensions for CF, proposing a new CF solution named \textbf{Conv}olutional \textbf{N}eural \textbf{C}ollaborative \textbf{F}iltering~(ConvNCF). 
As illustrated in Figure~\ref{fig:newframe},
ConvNCF has two characteristics making it distinct from existing models: 1) above user embedding and item embedding, we employ outer product (rather than concatenation or inner product) so as to explicitly capture the pairwise correlations between embedding dimensions, and 2) above the matrix generated by outer product, we employ convolution neural network (CNN) so as to learn high-order correlations in a hierarchical way. As ConvNCF concerns only the design of the predictive function, it is a general framework that can be specified with any embedding function that results in user embedding and item embedding. To show this universality, we devise three instantiations of ConvNCF by using the embedding function of three classical CF models --- 1) MF~\cite{rendle2009bpr}, which projects a user's ID into user embedding, 2) FISM~\cite{kabbur2013fism}, which projects a user's interacted items into user embedding, and 3) SVD++~\cite{koren2008factorization}, which projects a user's ID and interacted items into user embedding. We term the three specific methods as ConvNCF-MF, ConvNCF-FISM, and ConvNCF-SVD++, respectively, and conduct experiments on two real-world datasets to explore their effectiveness. Comparative results show that our ConvNCF methods outperform state-of-the-art CF methods in item recommendation, and extensive ablation studies verify the usefulness of both outer product and CNN in modeling embedding dimension correlations. To facilitate the research community, we have released the codes in: \href{https://github.com/duxy-me/ConvNCF}{https://github.com/duxy-me/ConvNCF}.

%Specifically, we devise an \textit{interaction map} to explicitly express the correlations, where the interaction map is particularly the outer product of the embeddings. In this way, all the interactions among the embedding dimensions are mapped to the elements in the interaction map. It is notable that the models using this interaction map subsumes the models using element-wise product, since the diagonal elements of interaction map are exactly the results of element-wise production. Due to the more substantial input features, to alleviate the effect of over-fitting, we leverage CNN, which is known to generalize better and is more easily to go deep than the fully connected MLP, as the interaction function to learn the inherent rules. To justify our proposal, we present three instantiations of ConvNCF by using different inputs to represent a user and conduct experiments on two real-world datasets. The extensive results verify the utility of modeling embedding dimension correlations with ConvNCF, which outperforms several competitive CF methods. Thus we believe that modeling embedding dimension correlations is necessary and universally applicable.

Note that a preliminary version of this work has been published as a conference paper in IJCAI 2018~\cite{he2018outer}. We summarize the main changes as follows:
\begin{enumerate}[leftmargin=*]
\item Introduction~(Section~\ref{sec:intro}). We reconstruct the abstract and introduction to emphasize the motivation of this extended version.

\item Methods~(Section~\ref{sec:framework} and Section~\ref{sec:methods}). We present ConvNCF as a general CF framework, and specify three methods that differ in the user embedding function to demonstrate its universality. The preliminary version only presents one specific method. 

\item Experiments~(Section~\ref{sec:expr}). This section is complemented with results of the two additional methods --- ConvNCF-FISM and ConvNCF-SVD++ --- to further justify the effectiveness of our proposal. 
%conduct extra experiments to show the generalizing ability of ConvNCF. Specifically, we add a number of extra comparisons between original CF models with the instantiations of ConvNCF, and deepen the explorations on training process.
\item Preliminaries (Section~\ref{sec:pre}) and Related Work~(Section~\ref{sec:related}). The two sections are newly added to make the paper more complete and self-contained. 
%We exhibit the recent developments on corresponding techniques, including collaborative filtering, matrix factorization, neural collaborative filtering, and convolutional neural networks, to depict the necessary background knowledge.
\end{enumerate}

%A preliminary version of this work has been published as a conference paper in IJCAI 2018~\cite{he2018outer}. This paper is significantly different from its preliminary version in the explorations on model generalization ability. Specifically, this work redefines the ConvNCF framework and proposes three instantiations. In order to verify the power of ConvNCF, we conduct a number of extra comparisons between original CF models with the instantiations of ConvNCF. We further enrich the explorations on training process. 

%The key contributions of this paper are as follows. 
%\begin{itemize}[leftmargin=*]
%	\item We breaks the inherent limitation of element-wise product and explicitly models embedding dimension correlations. 
%	\item We propose a novel neural network named ConvNCF for Cf tasks, which leverages CNN to learn high-order correlations among embedding dimensions from locally to globally in a hierarchical way. 
%	\item We conduct extensive experiments applying the representative embeddings on two public implicit feedback data, which demonstrate the effectiveness and rationality of ConvNCF methods. 
%	\item This is the first work that uses CNN to learn the correlations between user embedding and item embedding. It opens new doors of exploring the advanced and firstly evolving CNN methods for recommendation research.  
%\end{itemize}

The main contributions of this paper are as follows. 
\begin{itemize}[leftmargin=*]
	\item We propose a neural network framework named ConvNCF for CF, which explicitly models embedding dimension correlations and uses CNN to learn high-order correlations from locally to globally in a hierarchical way. 
	\item We implement three instantiations of ConvNCF that use different embedding functions to demonstrate the universality and effectiveness of ConvNCF. 
	%\item We conduct extensive experiments applying the representative embeddings on two public implicit feedback data. The results that ConvNCF methods outperform state-of-the-art CF methods demonstrate the effectiveness and rationality of ConvNCF methods. 
	\item This is the first work that explores the utility of capturing the correlations among embedding dimensions, providing a new path to improve recommendation models. 
%	uses CNN to learn the correlations between user embedding and item embedding. It opens new doors of exploring the advanced and firstly evolving CNN methods for recommendation research.  
\end{itemize}

\section{Related Work}
\label{sec:related}
This work lies in the topic of neural collaborative filtering. In this section, we first give a review of collaborative filtering on the embedding and interaction function, and then introduce the neural network based collaborative filtering. At last, to deepen the comprehension over neural network, we demonstrate the latest CNN developments and its applications in recommender systems.

\subsection{Collaborative Filtering}
User behaviors on the online platforms~(\ie purchasing, browsing or commenting) imply user preferences. By building the user-item interactions through these behavior records, Collaborative Filtering~(CF) is able to mine user hidden preferences. According to the types of user behaviors, CF can  be usually classified into two categories, namely explicit and implicit feedback based CF. Explicit feedback data such as user ratings directly indicates user's active evaluation. It has been a significant research task in the last decades.
To estimate the accurate score toward a specific item, various factorization models with a regression loss have been proposed. Particularly, models such as SVD++~\cite{koren2008factorization}, Localized MF~\cite{zhang2013localized}, Hierarchical MF~\cite{Suhang2015}, Social-aware MF~\cite{zhao2016user}, and CrossPlatform MF~\cite{CaoTOIS2017} have gained great success on specific tasks due to the ability to model specific contextual features. % One type of interactions is explicit feedback that 
% directly indicates user's active evaluations~(\eg user ratings) and preferences. 
% On this basis, the CF models, such as SVD++~\cite{koren2008factorization}, Localized MF~\cite{zhang2013localized}, Hierarchical MF~\cite{Suhang2015}, Social-aware MF~\cite{zhao2016user}, and CrossPlatform MF~\cite{CaoTOIS2017}, learn to fit the annotated data with regression loss. 
 Another type of interactions is known as implicit feedback, which records any user behaviors (e.g., what they watch and what item they buy). Implicit feedback data is more prevalent in practice since
 it does not require the user to express his taste explicitly~\cite{rendle2009bpr}. Hence, in this paper we aim to build recommendation algorithms based on implicit user feedback.
%  Due to the lower cost to collect, implicit feedback data is much more available 
%  Unlike rating data that requires user to express his taste explicitly, implicit data is much easier to be collected.
% is the most usual data widely existed in log files. In other words, collecting implicit data is much easier.
% Besides, implicit data is led by user interests though they are not the exact user preferences. Mining user preferences from it is now a popular CF task.

In most practical recommender systems, users usually only focus on the top ranked items rather than all rating scores. 
%Whether an interaction of a user and an item is observed in implicit data is not seriously related to whether the user likes the item. 
From this perspective, CF with implicit feedback seems like a personalized ranking problem rather than a score predicting problem. To address this problem, BPR~\cite{rendle2009bpr} loss is proposed to model the relative preferences between a pair of interactions, one of which is observed while the other is unobserved. The predicting score of the observed interaction must be higher than that of the unobserved one. Through this ranking scheme, BPR successfully trains a number of CF models~\cite{cao2017embedding,ACF},
%Moreover, BPR is a generic loss function, which is able to learn a wide range of recommendation models, 
including both shallow factorization models~\cite{rendle2009bpr} and deep neural network model~\cite{ACF}.
%explicit feedback also work over implicit feedback and achieve well performances with BPR loss. 
In fact, BPR is currently the dominant loss for CF models and has many improvements, such as improvedBPR~\cite{www2018improvedBPR,yuan2016lambdafm} by changing  the negative sampling to select informative examples and APR~\cite{he2018adversarial} by applying adversarial learning to enhance the model robustness.

\subsection{Matrix Factorization}
Matrix Factorization~(MF) is a significant technique in many domains~\cite{luo2016nonnegative,ma2018variational} due to its ability to distill co-occurrence patterns~\cite{ma2018decorrelation}. 
Within the variant CF models, MF is also an important class. Traditional MF algorithms work by decomposing the user-item interaction matrix into inner product of two lower-rank matrices~\cite{koren2009matrix}. Recently, the models representing users and items with two lower-rank matrices and predicting the interactions by inner product are considered to be MF family. Its subsequent works mainly focus on devising the user embedding and item embedding. 

The earliest MF model is FUNK-SVD\footnote{http://sifter.org/~simon/journal/20061211.html} that assigns a latent vector to each of the users and items. The prediction of FUNK-SVD is the inner product of the vectors. Despite the name, FUNK-SVD apply no singular value decomposition to get the model but use learning-based approach. Thus the embedding is easily extended with more complex structures. FISM~\cite{kabbur2013fism} combines the set of features from interacted items as user embedding. NAIS~\cite{he2018nais} weights the addition during combining process. 
SVD++~\cite{koren2008factorization} synthesizes latent vectors and the item correlations to construct hybrid embedding. DeepMF~\cite{DeepMF} applies MLP~\cite{gardner1998artificial} over the original embedding to abstract a high-level embedding. Additionally, the embedding could be generated with the content and context features~\cite{qian2019spatiotemporal,cheng2019mmalfm,wu2019context}. VBPR~\cite{he2016vbpr} takes latent vector composed with image features extracted by AlexNet~\cite{krizhevsky2012imagenet} as item embedding. CCF~\cite{lu2015content} proposes a content-based CF for news topic. Music recommendation~\cite{cheng2017exploiting} integrates the content-based feature to express the song. Some works even incorporate external knowledge~\cite{chen2018adversarial,pan2019transfer}. It is notable that some of these models were proposed for explicit feedback but they also perform well for implicit feedback by training with BPR loss~\cite{rendle2009bpr}.

\subsection{Neural Collaborative Filtering}
Neural network is known as a powerful data-based model well-performed in a wide range of domains. It is always described as a pipeline of layers. To meet the characteristics of different tasks, various novel structures of the neural network are proposed recently~\cite{he2017neural,qu2019product,feng2019temporal}. Multi-layer Perceptron~(MLP)~\cite{gardner1998artificial} is the fundamental neural network, the main body of which is composed of fully-connected layers, that outputs a projection of input features. Convolutional nerual network~(CNN)~\cite{krizhevsky2012imagenet,he2016deep} reduces the number of parameters of MLP and increases the number of layers with convolution operations, which automatically extract partial projection with convolution kernel. Attention mechanism~\cite{guan2019attentive,guo2019attentive} adapts the feature weights based on some auxiliary information, in order to capture more effective features. Due to the extraordinary performance on images, CNN has become the dominant 
module in the domain of computer vision.
%is now regarded as a core module in most computer vision tasks, 
% such as autoencoder~(AE)~\cite{kingma2013auto} and generative adversarial networks~(GAN)~\cite{goodfellow2014generative} for data generation and region-based convolutional Neural Networks~(RCNN)~\cite{girshick2015fast} for semantic segmentation.
More than that, both MLP and CNN models are widely used in the
 natural language processing~\cite{hochreiter1997long} and recommender system~\cite{he2017neural} domains.

Neural collaborative filtering~(NCF) is a family of models using neural network for the CF task. Most NCF methods focus on devising powerful interaction functions above the embedding layer instead of using the simple inner product. A popular substitute of inner product is to improve the element-wise product, such as weighted element-wise product~\cite{he2017neural}, or a MLP over the element-wise product~\cite{zhang2017joint}.

In the NCF task, MLP~\cite{gardner1998artificial}  has been extensively investigated in recent years, due to its simplicity and effectiveness. The basic idea of MLP is to stack multiple fully-connected layers, each of which are usually followed by a non-linear activation layer. % MLP~\cite{gardner1998artificial} is a neural network with multiple fully-connected layers. The fully-connected layer is a projected followed by a non-linear activation. Due to its simplicity, MLP is used in every fundamental network.
For example, NeuMF~\cite{he2017neural} devises a two-path model that one of the path is a MLP on top of the concatenation of user embedding and item embedding, and the other is a MLP above element-wise product of the embeddings. JRL~\cite{zhang2017joint} takes a MLP over the element-wised product of embeddings as input. Similarly, deepMF~\cite{DeepMF} learns a high-level embedding with MLP. CDAE~\cite{wu2016collaborative} on the other way devises an auto-encoder with fully-connected layers. Despite the success of MLP, its deficiency cannot be ignored. The models with MLP are easy to overfit and usually need more computing resources due to the large amount of parameters.

\subsection{Convolutional Neural Network}
In this work, we mainly investigate the utility of convolutional neural network~(CNN) in the CF task.
% Since convolutional neural network~(CNN) exhibits its power in image classification~\cite{krizhevsky2012imagenet}, it is used in most of the popular tasks~\cite{wang2012end,ronneberger2015u}.
The core of CNN is known as a partial feature extractor~(\ie convolutional layer). In each convolution layer, the output feature is not a projection of the whole input layer but several maps composed of partial features generated by convolution kernels. The key to obtain effective features with fewer parameters is that the partial features in the same map share a group of convolution kernels. Thus, compared with the fully-connected layer used in MLP, convolutional layer has much fewer parameters, which leads to more robust learning process and takes up fewer computing resources. 
% Thus it is widely accepted that CNN alleviates the overfitting of MLP.
%With this characteristic, CNN~\cite{huang2017densely} achieves the state-of-the-art dataset in the classic datasets MNIST~\cite{deng2012mnist}, Cifar~\cite{krizhevsky2010convolutional} and ImageNet~\cite{deng2009imagenet}. The idea extracting partial features makes CNN widely used in text~\cite{wang2012end}, audio~\cite{noda2015audio}, video~\cite{xu2015discriminative}, etc.
%Recent works replects many novel improvements over the basic CNN. ResNet~\cite{he2016deep} is proposed to recover the vanished gradient in very deep network. MobileNet~\cite{howard2017mobilenets} splits the convolutional kernel to achieve a higher efficiency. DenseNet~\cite{huang2017densely} leverages dense connections to obtain better performance. Attentive mechanism~\cite{neumann2017attentive} is proposed to make CNN focus on the key regions. Gated CNN~\cite{dauphin2016language} is a novel application in text processing. 
Due to these advantages, we believe CNN has the potential to substitute the MLP in the CF task with less computing cost and better recommendation accuracy.

In fact, several works that integrate CNN within the recommendation models have achieved success. The key idea of these works is to improve the original methods by feeding the recommemnder system with additional CNN features. One typical work is VBPR~\cite{he2016vbpr} that leverages the image features extracted with Alexnet~\cite{krizhevsky2012imagenet} to express the items more accurately. Similarly, ConvMF~\cite{kim2016convolutional} generates document latent features with a textual CNN. Though these works can provide more accurate auxiliary information, they rarely change the core pattern in predicting preference from the model perspective. 
NextItNet~\cite{yuan2019simple} is a newly proposed CNN model for the session-based recommendation. 
It combines masked filters with 1D dilated convolutions~\cite{yu2015multi} to increase the receptive fields when modeling long-range session data. Since this work targets at building a generic recommendation framework for traditional CF scenarios, we omit the detailed discussion of NextItNet.

\section{Preliminaries}
\label{sec:pre}
This section presents some technical preliminaries to the work. We first introduce the problem formulation of recommendation, and then discuss the recently proposed neural collaborative filtering framework~\cite{he2017neural}. \dxy{For the ease of reading, we use bold uppercase letter (\eg $\textbf{P}$) to denote a matrix, bold lowercase letter to denote a vector (\eg $\textbf{p}_u$), and calligraphic uppercase letter to denote a tensor (\eg $\mathcal{E}$).
Moreover, scalar $p_{u,k}$ denotes the $(u,k)$-th element of matrix $\textbf{P}$, and vector $\textbf{p}_u$ denotes the $u$-th row vector in $\textbf{P}$.
Let $\mathcal{E}$ be 3D tensor, then scalar $e_{a,b,c}$ denotes the $(a,b,c)$-th element of tensor $\mathcal{E}$, and vector $\textbf{e}_{a,b}$ denotes the slice of $\mathcal{E}$ at the element $(a,b)$. Table~\ref{tab:notations} and Table~\ref{tab:abbr} summarize the brief
parameters and abbreviations used in this paper, respectively.

\begin{table}[hbt]
    \centering
    \caption{Notations used in This Paper}
    \begin{tabular}{|l|l|}
        \hline
         Symbol & Definition \\\hline
         $u$, $i$ & the ids for user and item, respectively    \\
         $M$,$N$ & the numbers of users and items, respectively\\
         $K$ & the embedding size\\
         $\textbf{Y}$ & the observed user-item interaction matrix \\
         $y_{ui}$ & the observed interaction between user $u$ and item $i$ \\
         $\hat{y}_{ui}$ & the predicted interaction between user $u$ and item $i$ \\ 
         $f^U(u)$ & the function to capture the embedding of user $u$ \\
         $f^I(i)$ & the function to capture the embedding of item $i$ \\
         $\mathcal{F}(f^U(u),f^I(i))$ & the function to merge the input embedding $f^U(u)$ and $f^I(i)$\\
         
         \hline 
         
         $\textbf{E}$ & the interaction map \\
         $\textbf{E}^{lc}$ & the $c$-th feature map in convolutional layer $l$\\
         $\mathcal{E}^l$ & the 3D feature map of Layer $l$ \\
         $\mathcal{T}^{l}$ & the 4D convolutional kernel for Layer $l$ \\
         $b_l$ & the bias term for Layer $l$ \\
         $\textbf{w}$ & the weight vector for the prediction layer \\
         
         \hline 
         
         $\Theta_*$ & the trainable parameters \\
         $\lambda_*$ & the hyper parameters for regularizations \\
        
         \hline 
         
         $\textbf{P}$ & the user embedding matrix\\
         $\textbf{Q}$ & the item embedding matrix\\
         $\mathcal{R}_u$ & the set of items interacted by user $u$\\

         \hline
         
    \end{tabular}
    \label{tab:notations}
\end{table}

\begin{table}[hbt]
    \centering
    \caption{Abbreviations used in This Paper}
    \begin{tabular}{|l|l|}
        \hline
         Abbr. & Meaning \\\hline
         MF & Matrix Factorization \\
         FM & Factorization Machine \\\hline
         CF & Collaborative Filtering \\
         NCF & Neural Collaborative Filtering \\
         ConvNCF & Convolutional Neural Collaborative Filtering \\\hline
         MLP & Multi-Layer Perceptron \\
         CNN & Convolutional Neural Network \\

         \hline
         
    \end{tabular}
    \label{tab:abbr}
\end{table}
}

\subsection{Problem Formulation}
\label{subsec:cftask}

Recommendation is a large-scale personalized ranking task, which generates a distinct ranking list for each user. In model-based CF methods, the items are ranked by a predictive model $\hat{y}_{ui}$, which estimates how likely the user $u$ will consume the item $i$. That is, $\hat{y}_{ui}$ gives a high score if the user $u$ is likely to consume the item $i$, otherwise it gives a low score.

The predictive function $\hat{y}_{ui}$ is trained by learning parameters on the observed user-item interactions. Let $M$ and $N$ denote the number of users and items respectively, the observed user-item interaction data is represented as a matrix $\textbf{Y}\in\mathbb{R}^{M\times N}$. The element $y_{ui}$ located at $u$-th row and $i$-th column denotes the preference of user $u$ on item $i$. In most real-world scenarios, the interaction data is implicit feedback that records user behaviours, such as clicking, browsing, purchasing, etc. In this case, $y_{ui}$ is usually expressed as a binary variable to denote the user preference:
\begin{equation}
\label{eq:yui}
    y_{ui} = 
\begin{cases}
    1 & \text{if } (u,i) \text{ is observed,}\\
    0 & \text{otherwise.}
\end{cases}
\end{equation}

%Let $\hat{y}_{ui}$ be value of the prediction of $\mathcal{Y}(u,i)$ for short. 
%CF model aims to enture that, $\hat{y}_{ui} > \hat{y}_{ui}, \forall y_{ui} > y_{uj}$. The corresponding optimizing method is introduced in Section~\ref{subsec:training}. 
After the model is trained to fit the interaction data well, during the testing phase that recommends items to user $u$, we evaluate $\hat{y}_{ui}$ on all candidate items (e.g., items that are not interacted by the user or promotion items), and rank the items based on the values. 
%we firstly predict the user preference on all the $N$ items with well trained model $\mathcal{Y}(\cdot)$. Then from the scores $\mathcal{S} = \{s_1, s_2, ..., s_M\}$, we select top-$k$ score $\{s_{b_1}, s_{b_2}, ..., s_{b_k}\}$ and the corresponding items $\{b_1, b_2, ..., b_k\}$ are the items recommended for user $u$.

\subsection{Neural Collaborative Filtering}
\label{subsec:ncf}
% collaborative filterin focus on representation learning and predictive function. The key problem on predictive function is the absence of feature dimensions correlations. 

% this may should be in section NCF
There are two key ingredients in developing a CF model: 1) how to represent a user and an item, and 2) how to build the predictive function based on user and item representations.
In most cases, the users and items are represented as embeddings. Since we will present common embedding methods for CF in Section~\ref{sec:methods}, here we discuss more on the predictive function.  
%Since there are many classic embedding designs~\cite{he2016vbpr,kabbur2013fism,koren2008factorization}, here we do not discuss the design of embedding but explore the predictive function. 

%We just let $f^U(u)$ indicate user $u$ and $f^I(i)$ indicate item $i$, where $f^U(\cdot)$ and $f^I(\cdot)$ indicate the function converting ids to embeddings. 
%With the predefined embeddings, we explore how to better build the predictive function $\mathcal{Y}(u,i)$.

Neural collaborative filtering~(NCF)~\cite{he2017neural} proposes to parameterize the predictive function with feed-forward neural networks. Owing to the strong representation ability of neural networks in theory, NCF has been recognized as an effective approach to model user-item interactions and become a prevalent choice~\cite{ACF,zhang2017joint,DeepMF}.
%becomes a prevalent approach Referring to the definitions in the neural network, we call the operations in NCF as a layer, and organize them as a pipeline. Through the pipeline, data is transmitted from lower layer to higher layer. 
The predictive function in NCF can be abstracted as follows~\ref{eq:nn}:
\begin{equation}
    \label{eq:nn}
    \begin{aligned}
        \phi_0 &= \mathcal{F}(f^U(u),f^I(i)), \\
        \phi_l &= \mathcal{L}_l(\phi_{l-1}), \quad \text{for}\quad l = 1...L \\
        \hat{y}_{ui} &= \textbf{w}^T \phi_L, \\
    \end{aligned}
\end{equation}
where $f^U(\cdot)$ and $f^I(\cdot)$ denote the embedding function for users and items, respectively; $\mathcal{F}(\cdot)$ denotes the merge function that combines user embedding and item embedding to feed to hidden layers; $\mathcal{L}_l$ denotes the transformation function of the $l-$th hidden layer, $\phi_{l}$ denotes the output of the $l-$th layer; $\phi_L$ denotes the output of the last layer, and vector $\textbf{w}$ is to project $\phi_L$ to the final prediction score. Note that NCF presents a general framework such that each module in the framework is subjected to design. 
%It is notable that the type and property of neural network are determined by the implementation of $\mathcal{L}_*$, which we will discuss in Section~\ref{subsubsec:conv_mlp}.
%Multi-Layer Perceptron~(MLP) indicates that all the layers are fully-connected layers. Convolutional 

%To apply the user and item information to neural network, it is required to combine the user embedding and item embedding with $\mathcal{F}(\cdot)$, as the $\phi_0$ in Equation~\ref{eq:nn}. The recent two major definitions of $\mathcal{F}(\cdot)$ are element-wise product and concatenation. 

Next, we concentrate on the merge function $\mathcal{F}(\cdot)$ used in existing methods. 
%这里多引点工作
The two most prevalent choices in recommendation literature~\cite{DeepMF,zhang2017joint,ACF,he2017neural,cheng20183ncf,bai2017neural} are element-wise product and concatenation. 

\textbf{Element-wise Product} generates the combined vector by multiplying the corresponding dimension of user embedding and item embedding, that is, 
\begin{equation}
\label{eq:ewp}
\mathcal{F}(f^U(u),f^I(i)) = 
\left(
\begin{matrix}
    f^U(u)_1\cdot f^I(i)_1,\\
    f^U(u)_2\cdot f^I(i)_2,\\
    \cdots, \\
    f^U(u)_K\cdot f^I(i)_K\\
\end{matrix}
\right)
\end{equation}
The rationality of this operation for CF stems from inner product --- by summing the elements in the resultant vector, it is equivalent to the output of inner product. 
Note that inner product is an effective approach to measure the vector similarity, having been widely used in classic CF models~\cite{kabbur2013fism,koren2008factorization}. As such, many neural network take it for granted to use element-wise product, such as JRL~\cite{zhang2017joint} and NeuMF~\cite{he2017neural}. 
%However, after element-wise product, the raw information is lost.

\textbf{Concatenation} appends the elements in item embedding to the user embedding vector: 
\begin{equation}
\label{eq:cc}
\mathcal{F}(f^U(u),f^I(i)) = 
[f^U(u)_1, ..., f^U(u)_K, f^I(i)_1, ..., f^I(i)_K]^T
\end{equation}
This operation keeps the raw information in user and item embeddings, without explicitly modeling any interaction between them.  
As such, the model has to expect the following hidden layers to learn the interaction signal in the embeddings.
%However, due to the strong expressiveness of NN models and the insufficiency of data, learning the rules from such data are always difficult. It is observed that the MLP~\cite{gardner1998artificial} CF model performs a bit worse than JRL~\cite{zhang2017joint}.

It is worth highlighting that one main drawback of the two merge functions is that, the embedding dimensions are assumed to be independent. In other words, no correlations among the dimensions are captured --- in element-wise product only the interactions of the same dimension are modeled, whereas in concatenation no any interactions are considered. 
Although the following hidden layers might be able to learn the correlations, such process is rather implicit and there is no guarantee that the desired correlations can be successfully captured. One empirical evidence is that the MLP (i.e., concatenation followed by three fully connected hidden layers) even underperforms the simple MF model~\cite{he2017neural}. Another evidence is from ~\cite{tay2018latent}, which shows that MLP has to use much more parameters to approximate the inner product well, indicating the limitation in representation ability.

%approximate any continuous function~\cite{cybenko1989approximation}, therefore it might be capable of capturing  the  possible  correlations. However,  such  process is rather implicit and as such, it may be ineffective in capturing certain  relations -- an  evidence  is  from showing  that much more parameters have to be used in order to approximate the simple multiplicative relation.

\section{ConvNCF Framework}
\label{sec:framework}
This section presents our proposed ConvNCF framework. We first give an overview of the framework, followed by elaborating the key design of outer product and convolutional layers for modeling the embedding dimension correlations. Finally we describe the model prediction and training method.  
%In this section, we first give an overview of \textbf{Conv}olutional \textbf{N}eural \textbf{C}ollaborative \textbf{F}iltering~(ConvNCF) framework. Then we elaborate the core layers, outer product layer and convolutional layers. At last we describe the training and predicting processes of ConvNCF. 

\subsection{Framework Overview}
\dxy{We focus on the neural structures since the neural network-based models in recent works always perform well in various learning models~\cite{huang2017densely,he2017neural,ACF}. Then we select CNN as our fundamental neural structure because, 1) the feature map performs as a 2D matrix, which naturally adapts to the input of CNN, 2) the subregion in the feature map has a spatial relationship, \ie the correlation among multiple dimensions, which could be represented by convolutional filter, and 3) through the convolutional layers, the correlations among all embedding dimensions could be captured from locally to globally. Compared with MLP, the usual neural structure in neural recommendation models, CNN has fewer parameters which lead to less over-fitting. Therefore, we propose the CNN-based framework, ConvNCF.
}

Figure~\ref{fig:newframe} illustrates the ConvNCF framework, with the embedding size of 64 and six convolution layers as an example. The embedding layer (left most) contains the embedding functions $f^U(u)$ and $f^I(i)$, which outputs two vectors  (of size 64) to represent user $u$ and item $i$, respectively. 
%The left most of the framework is the embedding layer, which convert user $u$ and item $i$ to their embeddings $f^U(u) \in \mathbb{R}^{64}$ and $f^I(i)\in \mathbb{R}^{64}$. 
Above the embedding layer, ConvNCF models the pairwise correlations between embedding dimensions by constructing the \textit{Interaction Map} $\textbf{E}\in \mathbb{R}^{64\times 64}$, which is the \textbf{outer product} of user embedding and item embedding. The interaction map is then fed to a stack of \textbf{convolutional layers} to learn high-order correlations; the convolutional layers follow a tower structure with 32 feature maps in the example, and the last convolutional layer outputs a tensor sized $1\times 1 \times 32$. In the prediction layer, ConvNCF applies a linear projection on the $1\times 1 \times 32$ tensor to obtain the prediction $\hat{y}_{ui}$. Given the model prediction, ConvNCF is trained with the pairwise learning objective BPR (Bayesian Personalized Ranking~\cite{rendle2009bpr}).

%Then we describe the outer product layer, the convolutional layers and the prediction layer in following text and leave the description on embedding types in Section~\ref{sec:methods}. 

\subsection{Outer Product Layer}
We merge user embedding $f^U(u)\in \mathbb{R}^{K\times 1}$ and item embedding $f^I(i)\in \mathbb{R}^{K\times 1}$ with an outer product operation, which results in a matrix $\textbf{E}\in \mathbb{R}^{K \times K}$: 
%Outer product layer is a function of $\mathcal{F}(\cdot)$ which combines the user embedding vector $f^U(u)$ and item embedding vector $f^I(i)$. The interaction map of $f^U(u)$ and $f^I(i)$ is denoted as
%$\textbf{E} \in \mathbb{R}^{K \times K}$  via the outer product operation:
% In this layer,  $f^U(u)$ and $f^I(i)$ are incorporated as interaction map $\textbf{E}$ via outer product,
\begin{equation}
\begin{aligned}
\label{eq:outer}
    \textbf{E} = f^U(u) \otimes f^I(i) = f^U(u) \cdot f^I(i)^T,
\end{aligned}
\end{equation}
where  the ($k_1,k_2$)-th element in $\textbf{E}$ is: $e_{k_1, k_2} = f^U(u)_{k_1}\cdot f^I(i)_{k_2}$. 
It can be seen that all pairwise embedding dimension correlations are encoded in the  $\textbf{E}$, thus we term it as \textit{Interaction Map}. 

Compared to the widely used inner product operation or element-wise product, we argue that interaction map generated via outer product is more advantageous in threefold: 1) it subsumes the element-wise product which considers only diagonal elements in our interaction map; 2) it encodes more signal by accounting for the correlations between different embedding dimensions; and 3) it is more meaningful than the simple concatenation operation, which only retains the original information in embeddings without modeling any correlation. 
Moreover, it has been recently shown that, modeling the interaction of feature embeddings explicitly is particularly useful for a deep learning model to generalize well on sparse data, whereas using concatenation is less effective~\cite{NFM,LatentCross}. 

Another potential benefit of the interaction map lies in its 2D matrix format --- which is the same as an image. In this respect, the pairwise correlations encoded in our interaction map can be seen as the local features of an ``image''. 
As we all know, deep learning methods have achieved the most success in computer vision, and many powerful deep models especially the ones based on CNN (e.g., 
%AlexNet~\cite{krizhevsky2012imagenet}, 
ResNet~\cite{he2016deep} and DenseNet~\cite{huang2017densely}) have been developed for learning from 2D image data. Thus, building a 2D interaction map allows these powerful CNN models to be applied in  the recommendation task.

\subsection{Convolutional Layers}
The interaction map \textbf{E} is fed into multiple convolutional layers for learning high-order correlations. As Figure \ref{fig:newframe} illustrates, there are 6 convolutional layers with the kernel size $2\times 2$ and stride 2, meaning that each successive layer is half size of the previous layer. Each convolutional layer  
%an illustrative example of our ConvNCF model. The embedding dimension of our model is usually set as  $2^N$. For example, when we use $64$-dimensional embedding, the interaction map becomes a $64\times 64$ matrix. Above the interaction map, there are 6 convolutional layers. 
is composed of 32 convolution kernels (note that pooling is not used in this work).
%Under this setting, there are four properties: 1) each element in the output denotes an $2\times 2$ area in the input, 2) all the $2\times 2$ areas covers the entire input precisely, 3) the width and height of the output tensor is half of that of the input tensor, and 4) the output of the sixth layer is $1\times 1\times 32$ where $32$ is the number of the convolutional kernels. 
In this structure, higher layers extract higher-order correlations among dimensions. For example, an element in Layer 1 indicates a $2\times 2$ area in $\textbf{E}$, while an element in Layer 2 indicates a $4\times 4$ area in $\textbf{E}$. The $4\times 4$ area indicates the correlations among 4 dimensions. Thus the last layer (i.e., the Layer 6 in Figure~\ref{fig:newframe}) contains the correlations among all the input embedding dimensions.

\dxy{There are two reasons that convolutional layers would be effective over the feature map. Firstly, as a local feature extractor, the convolutional filter does not require that all the subregions have the same rule. In image applications, the image patches cropped from the border and the center are much different. Similarly, in text applications, the phrases captured from the beginning, the middle, and the end of the sentences follow different grammar rules. Secondly, the interaction map may have a spatial relationship, where the subregion indicates the correlations between multiple dimensions, as discussed in Section 4.3.3.
}

%\dxy{Above the interaction map, there are many choices of layers which impact the performance. Recent works~\cite{he2017neural,ACF,huang2017densely} demonstrate that the neural network-based models always perform well not only in the recommender systems but also in various learning models. Thus, only the neural models are considered in our work. Moreover, compared with MLP, CNN is known as an equally powerful neural model with fewer parameters, which lead to less over-fitting. 
%Intuitively, the convolutional filter is just a partial feature extractor, but never demands that each subregion respects to the same distribution. For example, the image patches cropped from the border and the center are much different, and similarly, the phrases captured from the beginning, the middle, and the end of sentences follow different grammar rules. In fact, there is an equivalent conversation between the convolutional layer and the fully-connected layer\footnote{More details could be found in \url{http://cs231n.github.io/convolutional-networks/#convert}}.
%Therefore, we propose a novel CNN-based framework instead of the common choice of MLP in neural recommendation~\cite{he2017neural,zhang2017joint}. In what follows, we discuss the advantages and then give more details on how the convolution operation facilitates the implementation. }
%In the final layer, ConvNCF predicts $\Hat{y}_{ui}$ by integrating the last output values.

\subsubsection{Advantages over MLP} 
\label{subsubsec:conv_mlp}

Above the interaction map, the choice of neural layers has a large impact on its performance. A straightforward solution is to use the MLP network as proposed in NCF~\cite{he2017neural}; note that to apply MLP on the 2D interaction matrix $\textbf{E}\in \mathbb{R}^{K\times K}$, we can flat $\textbf{E}$ to a vector of size $K^2$.
Despite that MLP is theoretically guaranteed to have a strong representation ability~\cite{hornik1991approximation}, its main drawback of having a large number of parameters cannot be ignored. 
As an example, assuming we set the embedding size of a ConvNCF model as 64 (i.e., $K=64$) and follow the common practice of the half-size tower structure.
%~\cite{Covington:2016}. 
In this case, even a 1-layer MLP has $8,388,608$ (i.e., $4,096\times 2,048$) parameters, not to mention the use of more layers. We argue that such a large number of parameters makes MLP prohibitive to be used for prediction because of three reasons: 1) It requires powerful machines with large memories to store the model; and 2) It needs a large number of training data to learn the model well; and 3) It needs to be carefully tuned on the regularization of each layer
%~(e.g., dropout~\cite{dropout})
to ensure good generalization performance\footnote{In fact, another empirical evidence is that most papers used MLP with at most 3 hidden layers, and the performance only improves slightly (or even degrades) with more layers~\cite{he2017neural,Covington:2016,NFM}}.

In contrast, convolution filter can be seen as the ``locally connected weight matrix'' for a layer, since it is shared in generating all entries of the feature maps of the layer. This significantly reduces the number of parameters of a convolutional layer compared to that of a fully connected layer. Specifically, in contrast to the 1-layer MLP that has over 8 millions parameters, the above 6-layer CNN has only about 20 thousands parameters, which are several magnitudes smaller. This not only allows us to build deeper models than MLP easily, but also benefits the stable and generalizable learning of high-order correlations among embedding dimensions.

\subsubsection{Details of Convolution}
%In order to facilitate understanding of the details of convolution, we formulate the process of CNN stacks.
%First, in Figure \ref{fig:newframe}, CNN workflow starts from the interaction map $\textbf{E}$ with the size of $64\times 64$, and the model has 6 convolutional layers, where each convolutional layer has 32 convolutional kernels. Through each kernel, the layer would output one feature map. One convolutional layer would output 32 feature maps.
Each convolution layer has 32 kernels, each of which produces a feature map. A feature map $c$ in convolutional layer $l$ is represented as a 2D matrix $\textbf{E}^{lc}$; since we set the stride to 2, the size of $\textbf{E}^{lc}$ is half of its previous layer $l-1$, e.g., $\textbf{E}^{1c}\in \mathbb{R}^{32\times 32}$ and $\textbf{E}^{2c}\in \mathbb{R}^{16\times 16}$. 
For Layer $l$, all feature maps together  can be represented as a 3D tensor $\mathcal{E}^{l}$. 

Given the input interaction map $\textbf{E}$, we can first get the feature maps of Layer 1 as follows:
\begin{equation}\small
\label{eq:conv1}
\begin{aligned}
\mathcal{E}^{1} &= [e^{1}_{i,j,c}]_{32\times 32\times 32}, \quad \text{where} \\
e^{1}_{i,j,c} &= \text{ReLU}(b_1 + \sum_{a=0}^1 \sum_{b=0}^1
 e_{2i+a,2j+b}\cdot \underbrace{  t^{1}_{a,b,c}}_{\text{convolutional kernel}}),
\end{aligned}
\end{equation}
where $b_1$ denotes the bias term for Layer 1, and $\mathcal{T}^{1}=[t^{1}_{a,b,c}]_{2
\times 2\times 32}$ is a 3D tensor denoting the convolutional kernel for generating feature maps of Layer 1. We use the rectifer unit as activation function, a common choice in CNN to build deep models. 
Following the similar convolution operation, we can get the feature maps for the following layers. The only difference is that from Layer 1 on, the input to the next layer $l+1$ becomes a 3D tensor $\mathcal{E}^l$:
\begin{equation}\small
\label{eq:conv2}
\begin{aligned}
\mathcal{E}^{l+1} &= [e^{l+1}_{i,j,c}]_{s\times s\times 32}, \quad \text{where}\  1\leq l\leq 5, \ s = \frac{64}{2^{l+1}}, \\ e^{l+1}_{i,j,c} &= \text{ReLU}(b_{l+1} + \sum_{a=0}^1 \sum_{b=0}^1
 \textbf{e}^l_{2i+a,2j+b}\cdot \textbf{t}^{l+1}_{a,b,c}),
\end{aligned}
\end{equation}
where $b_{l+1}$ denotes the bias term for Layer $l+1$, and  $\mathcal{T}^{l+1}=[t^{l+1}_{a,b,c,d}]_{2\times 2\times 32\times 32}$ denote the 4D convolutional kernel for Layer $l+1$. The output of the last layer is a tensor of dimension $1\times 1 \times 32$, which can be seen as a vector and is projected to the final prediction score with a weight vector $\textbf{w}$. 

%In the interaction map $\textbf{E}$, each entry $e_{ij}$ encodes the second-order correlation between the dimension $i$ and $j$. Next, each hidden layer $l$ captures the correlations of a $2\times 2$ local area\footnote{The size of the local area is determined by our setting of the filter size, which is subjected to change with different settings.} of its previous layer $l-1$. As an example, the entry $e^{1}_{x,y,c}$ in Layer 1 is dependent on four elements $[e_{2x,2y}; e_{2x,2y+1}; e_{2x+1,2y}; e_{2x+1, 2y+1}]$, which means that it captures the 4-order correlations among the embedding dimensions $[2x; 2x+1; 2y; 2y+1]$. Following the same reasoning process, each entry in hidden layer $l$ can be seen as capturing the correlations in a local area of size $2^{l}$ in the interaction map $\textbf{E}$. 

%As such, an entry in the last hidden layer encodes the correlations among all dimensions. Through this way of stacking multiple convolutional layers, we allow ConvNCF to learn high-order correlations among embedding dimensions from locally to globally, based on the 2D interaction map. 

\subsubsection{Dimension Correlation in ConvNCF}
\dxy{
Here we provide detailed explanations on how ConvNCF models the dimension correlations between embedding factors. Equation~\ref{eq:outer} demonstrates that the entry $e_{xy}$ in the interaction map $\textbf{E}$ is a product of $f^U(u)_x$ and $f^I(i)_y$, which is viewed as a 1-to-1 correlation, also namely 1-order correlation or basic correlation. Let $[xs:xe]$ be a row range and $[ys:ye]$ be a column range, the entries in the adjacent subregion $\textbf{E}_{xs:xe,ys:ye}$ indicate all the basic correlations between $f^U(u)_{xs:xe}$ and $f^I(i)_{ys:ye}$. Overall, the interaction map $\textbf{E}$ itself contains all the basic correlations between $f^U(u)$ and $f^I(i)$. Note that, the first layer of ConvNCF is the interaction map $\textbf{E}$, which means what ConvNCF actually does is to predict a score by modeling all the basic correlations.

Then we elaborate on how ConvNCF models the correlations. Equation~\ref{eq:conv1}  %v Equation~\ref{eq:conv2} 
demonstrates that the feature $\textbf{e}^1_{xy}$ is the composite correlation of four entries in the interaction map $\textbf{E}$, $[e_{2x,2y}; e_{2x,2y+1};$ $e_{2x+1,2y}; e_{2x+1, 2y+1}]$. Thus the entry $\textbf{e}^1_{xy}$ in the the feature map $\textbf{E}^1$ is actually the feature of the composite correlations of  $\textbf{E}_{2x:2x+1,2y:2y+1}$, which is a 2-to-2 correlation namely 2-order correlation. Therefore, the feature map $\textbf{E}^1$ is composed of 2-order correlations. Similarly, above the feature map $\textbf{E}^1$, Equation~\ref{eq:conv2} demonstrates that the entry $\textbf{e}^2_{xy}$ in $\textbf{E}^2$ is the composite correlations of $\textbf{E}^1_{2x:2x+1,2y:2y+1}$, which is the 4-to-4 composite correlations of $\textbf{E}_{4x:4x+3,4y:4y+3}$ namely 4-order correlation. Thus, the feature map $\textbf{E}^2$ is composed of 4-order correlations.
It is worth mentioning that the convolutional kernels are sized $2\times2$ with stride 2 and no padding. The entries in higher feature map can just cover all the entries in lower feature map. This ensures the integrity in transferring the embedding information. As such, an entry in the last hidden layer encodes the correlations among all dimensions. Through this way of stacking multiple convolutional layers, we allow ConvNCF to learn high-order correlations among embedding dimensions from locally to globally, based on the 2D interaction map. 
}

%Here we give some intuitions on how ConvNCF can capture high-order correlations among embedding dimensions. In the interaction map $\textbf{E}$, each entry $e_{ij}$ encodes the second-order correlation between the dimension $i$ and $j$. Next, each hidden layer $l$ captures the correlations of a $2\times 2$ local area\footnote{The size of the local area is determined by our setting of the filter size, which is subjected to change with different settings.} of its previous layer $l-1$. As an example, the entry $e^{1}_{x,y,c}$ in Layer 1 is dependent on four elements $[e_{2x,2y}; e_{2x,2y+1}; e_{2x+1,2y}; e_{2x+1, 2y+1}]$, which means that it captures the 4-order correlations among the embedding dimensions $[2x; 2x+1; 2y; 2y+1]$. Following the same reasoning process, each entry in hidden layer $l$ can be seen as capturing the correlations in a local area of size $2^{l}$ in the interaction map $\textbf{E}$. 
%As such, an entry in the last hidden layer encodes the correlations among all dimensions. Through this way of stacking multiple convolutional layers, we allow ConvNCF to learn high-order correlations among embedding dimensions from locally to globally, based on the 2D interaction map. 

\subsection{Model Prediction and Training}
\label{subsec:training}
Given the output vector of the last convolutional layer as $\textbf{g}\in \mathbb{R}^{K\times 1}$, the model prediction is defined as: $\hat{y}_{ui} = \textbf{w}^T\textbf{g}$, where $\textbf{w}$ is the trainable weight vector in the prediction layer. 
%The target of CF is to estimate $\hat{y}_{ui} = \mathcal{Y}(u,i)$. In ConvNCF, the prediction layer defines $\mathcal{Y}(u,i) = \textbf{w}^T\textbf{g}$, where $\textbf{g}$ is the flattened output of convolutional layers and $\textbf{w}$ is a projection vector. 
To summarize, the parameters in ConvNCF are grouped into four parts, $\Theta_U$ in user embedding function $f^{U}(\cdot)$, $\Theta_I$ in item embedding generation $f^{I}(\cdot)$, $\Theta_{CNN}$ in the convolutional layers, and $\textbf{w}$ for the prediction layer. 
%Thus, the prediction of ConvNCF is $\hat{y}_{ui} = \mathcal{Y}(u,i|\Theta_u, \Theta_i, \Theta_{cnn}, \textbf{w})$.

Considering that recommendation is a personalized ranking task, we train the parameters with a ranking-aware objective function. Here we adopt the Bayesian Personalized Ranking~(BPR) objective function~\cite{rendle2009bpr}:
\begin{equation}\small\label{eq:bprconv}
\begin{aligned}
    L &= \sum_{(u,i,j)\in \mathcal{D}} -\ln \sigma(\hat{y}_{ui} - \hat{y}_{uj})\\
    &+ \lambda_1 ||\Theta_U||^2 + \lambda_2 ||\Theta_I||^2 + \lambda_3 ||\Theta_{CNN}||^2 +  \lambda_4 ||\textbf{w}||^2 , 
\end{aligned}
\end{equation}
where $\lambda_*$ are hyper-parameters for regularization, and $\mathcal{D}$ denotes the set of training instances: $\mathcal{D}:= \{(u,i,j)| i\in \mathcal{R}_u^+ \wedge j\notin  \mathcal{R}_u^+ \}$, where $\mathcal{R}_u^+$ denotes the set of items that has been consumed by user $u$.

In each training epoch, we first shuffle all observed interactions, and then 
get a mini-batch in a sequential way. 
Given the mini-batch of observed interactions, we generate negative examples on the fly to get the training triplets. The negative examples are randomly sampled from a uniform distribution; while recent efforts show that a better negative sampler can further improve the performance~\cite{www2018improvedBPR,yuan2016lambdafm}, we leave this exploration as future work. 
\section{Three ConvNCF Methods}
\label{sec:methods}
\begin{figure}[t!]
	\centering
	\includegraphics[width=0.8\linewidth]{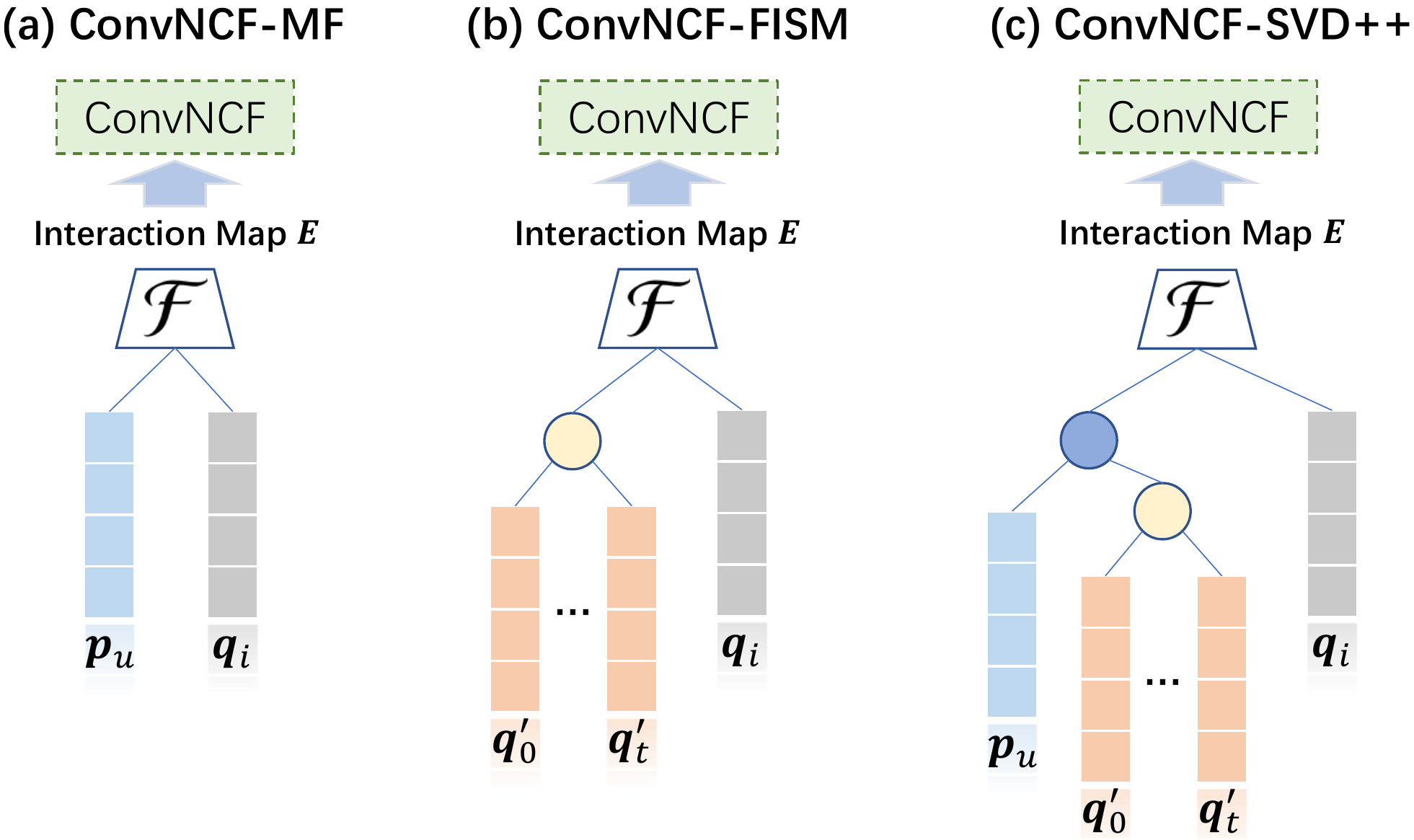}
	\caption{Illustration of the embedding function of the three ConvNCF methods.}
	\label{fig:embs}
\end{figure}

To demonstrate how ConvNCF works, we propose three instantiations of ConvNCF by equipping them with different embedding functions. We name the three methods as \textbf{ConvNCF-MF}, \textbf{ConvNCF-FISM} and \textbf{ConvNCF-SVD++}, which adopt the embedding function used in matrix factorization (MF)~\cite{rendle2009bpr}, factored item similarity model (FISM)~\cite{kabbur2013fism}, and SVD++~\cite{koren2008factorization}, respectively. Figure~\ref{fig:embs} illustrates the three embedding mechanisms, which differ in the user embedding function part (the item embedding functions are the same, which project item ID to embedding vector). \dxy{Then the embedding vectors are fed into the merge function $\mathcal{F}$ (\ie the outer product) to generate the interaction map $\textbf{E}$, which is fed into the ConvNCF subsequently.} Next, we present the three methods in turn; and lastly, we discuss the important tricks of embedding pre-training and adaptive regularization, which are crucial for the effectiveness of ConvNCF methods. 

\subsection{ConvNCF-MF}
%The three types of embedding are from classic CF models, with three kinds of concepts. According to the original model of the embeddings, we call the instantiations as \textbf{ConvNCF-MF}, \textbf{ConvNCF-FISM} and \textbf{ConvNCF-SVD++}. Regardless of the types, each of the users and items are represented by one vector, which is supplied to the function $\mathcal{F}$ for the subsequent processing, as shown in Figure~\ref{fig:embs}. Here we introduce the three types of embedding respectively. 

MF adopts the simplest ID embedding function, i.e., directly projecting the ID of a user (an item) to latent vector representation:
\begin{equation}
\label{eq:mf}
f^U(u) = \textbf{p}_u, \quad
f^I(i) = \textbf{q}_i.
\end{equation}
Let the user embedding matrix be $\textbf{P}\in\mathbb{R}^{M\times K}$ and the item embedding matrix be $\textbf{Q}\in\mathbb{R}^{N\times K}$. This function is equal to taking out the $u$-th row of $\textbf{P}$ as $u$'s embedding vector (i.e., $\textbf{p}_u$) and the $i$-th row of $\textbf{Q}$ as $i$'s embedding vector (i.e., $\textbf{q}_i$). 

In MF, each user is associated with an individualized set of parameters to denote the user's interest --- even two users have exactly the same interactions on items, they are parameterized differently in the model. As such, it is also called as \textit{user-based CF} in recommendation literature. The problem of this setting is that, if a user has never been seen in the training set (i.e., cold-start), the model cannot obtain her embedding in the online phase, even though the user has some new interactions on items. In other words, the model lacks the ability to provide real-time personalization. Next, we introduce the setting of \textit{item-based CF}, we can address this deficiency. 

%The original embedding for ConvNCF proposed in our conference version~\cite{he2018outer} is from MF-BPR~\cite{rendle2009bpr}, which is the fundamental CF model assigning latent features to each user and each item. Formally, let $K$, $M$, and $N$ denote the embedding size, number of user features, and number of item features, respectively, $\textbf{P}\in\mathbb{R}^{M\times K}$ denote the $M$ user embeddings and $\textbf{Q}\in\mathbb{R}^{N\times K}$ denote the $N$ item embeddings, the embedding for user $u$ and item $i$ are

%This embedding is known as the representative user-based embedding and it is most popular in CF due to its simplicity and effectiveness. There is another general definition of MF embedding. Let $\textbf{v}_u^U$ and $\textbf{v}_i^I$ be the feature vector for user $u$ and item $i$, respectively, we can obtain their embeddings $\textbf{p}_u$ and $\textbf{q}_i$ via

\subsection{ConvNCF-FISM}
Instead of directly projecting user ID to embedding vector, FISM represents a user with her historically interacted items, based on which the embedding function is defined:
%We select the representative technique FISM~\cite{kabbur2013fism} for our second instantiation.  FISM assigns a latent vector to each of the $N$ items, but it presents a user embedding with the items the user interacted with. Following FISM, let $\textbf{Q}^p \in \mathbb{R}^{N\times K}$ indicates the item factor for user representations and $\textbf{Q} \in \mathbb{R}^{N\times K}$ indicates the latent item embeddings, our feature encoder $f^U(\cdot)$ and $f^I(\cdot)$ are defined in Equation \ref{eq:fism}.
\begin{equation}
    \label{eq:fism}
    f^U(u) = \frac{1}{|\mathcal{R}_u|^\alpha}\sum_{t\in \mathcal{R}_u\backslash\{i\}} \textbf{q}'_t,\quad 
    f^I(i) = \textbf{q}_i, 
\end{equation}
where $\mathcal{R}_u$ denotes the set of items interacted by $u$, $\textbf{q}'_t$ denotes the embedding of historical item $t$ in constructing the user embedding function. Note that FISM has three considerations to ensure the embedding quality: 1) the historical item embedding $\textbf{q}'_t$ used in user embedding function is different from target item embedding $\textbf{q}_t$ used in item embedding function, which can improve the model representation ability; 2) the target item $i$ is excluded to represent the user, i.e., the $\mathcal{R}_u\backslash\{i\}$ term, so as to avoid information leakage in training; 3) the coefficient $\frac{1}{|\mathcal{R}_u|^\alpha}$ is to normalize users of different activity levels (by convention we set $\alpha$ as 0.5). 

In FISM, a user's embedding is aggregated from the embeddings of the user's historically interacted items. Thus, for two users with the same interactions on items, their embeddings are the same. For a cold-start user, as long as she has new interactions, we can obtain her embedding instantaneously by calling the embedding function without re-training the model. Therefore, such item-based CF scheme is suitable to provide real-time personalization to refresh the recommendation for users that have new interactions.  
%, and $\alpha$ is a hyper-parameter which is set to $0.5$ empirically in our experiments. 
%In order to avoid information leakage, we eliminate the embedding of item $i$ when generating the user embedding.

\subsection{ConvNCF-SVD++}
SVD++ is a hybrid method that combines the user embedding design of MF and FISM. It uses both the ID and historically interacted items to represent a user, and defines the embedding function as:
%The third instantiation of ConvNCF is based on a hybrid embedding SVD++~\cite{koren2008factorization}, which is also known as a powerful CF model. The model SVD++ wins the Netflix Prize competition~\cite{bennett2007netflix} and performs its power in most applications. Following SVD++, our hybrid feature extractor $f^U(\cdot)$ and $f^I(\cdot)$ are defined in Equation \ref{eq:svdpp}.
\begin{equation}
    \label{eq:svdpp}
    f^U(u) = \textbf{p}_u + \frac{1}{|\mathcal{R}_u|^\alpha}\sum_{t\in \mathcal{R}_u\backslash\{i\}} \textbf{q}'_t, \quad 
    f^I(i) = \textbf{q}_i, 
\end{equation}
where the notations follow the ones used in Equation~(\ref{eq:mf}) and (\ref{eq:fism}). The user embedding function of SVD++ is the sum of the user embedding functions of MF and FISM, unifing the strengths of both methods. Specifically, $\textbf{p}_u$ is a static vector to encode user inherent preference, and $\sum_{t\in \mathcal{R}_u\backslash\{i\}} \textbf{q}'_t$ can be dynamically adjusted by including recently interacted items. Note that SVD++ is a very competitive CF model that is known as the best single model in the Netflix Challenge~\cite{bennett2007netflix}. By plugging its embedding functions into ConvNCF, we can further advance its performance. 

%The definitions of the variables are the same as Equation \ref{eq:mf} and Equation \ref{eq:fism}. Similarly,  we eliminate the embedding of item $i$ when generating the user embedding and empirically set the hyper parameter $\alpha$ to $0.5$. This is a simple but effective integration of user-based and item-based models.

\subsection{Embedding Pre-training and Adaptive Regularization}
Due to the multi-layer design of ConvNCF, the initialization of model parameters has a large impact on model training and its testing performance. As the convolutional layers are learned on the outer product of embeddings to generate the features used for prediction, a good initialization on embeddings is beneficial to the overall model learning. Thus, we propose to pre-train the embedding layer. For the three ConvNCF methods, we first train its shallow model counterpart, i.e., MF, FISM, and SVD++ for ConvNCF-MF, ConvNCF-FISM, and ConvNCF-SVD++, respectively. We then use the learned embeddings to initialize the embedding layer of the corresponding method. Our empirical studies find that this strategy can substantially improve the model performance. 
%Embedding, the representation of users and items, consists of a large amount of parameters. It should be learned during the training process. In order to accelerate the training process and optimize the performance, we pre-train the embedding matrices with original methods.

After pre-training the embedding layer, we start training the ConvNCF model. As other model parameters are randomly initialized, the overall model is in an underfitting state. Thus, we disable regularization for the 1st epoch to make the model learn to fit the data as quickly as possible. For the following epochs, we enforce regularization on ConvNCF to prevent overfitting, including the $L_2$ regularization on the embedding layer (controlled by $\lambda_1$ and $\lambda_2$), convolution layers (controlled by $\lambda_3$), and the output layer (controlled by $\lambda_4$). It is worth noting that the regularization coefficients (especially $\lambda_4$ on the output layer) have a very large impact and should be carefully tuned for an optimal performance. 

\begin{table*}
\newcommand{\specialcell}[2][c]{%
  \begin{tabular}[#1]{@{}c@{}}#2\end{tabular}}
	\centering
	\caption{Top-$k$ recommendation performance where $k\in\{5,10,20\}$. RI indicates the average improvement of ConvNCF-SVD++ over the baseline. $^*$ indicates that the improvements over all baselines are statistically significant for $p<0.05$.}
	\resizebox{\textwidth}{!}{%
	\begin{tabular}{|c|c|l|l|l|l|l|l|l||c|}
		\hline
		
		\multirow{2}{*}{Dataset}&\multirow{2}{*}{Model Type}&\multirow{2}{*}{Model}& \multicolumn{3}{c|}{\textbf{HR@$k$}} & \multicolumn{3}{c||}{\textbf{NDCG@$k$}} & \multirow{2}{*}{RI}\\\cline{4-9}
		&&& $k=5$    & $k=10$   & $k=20$   & $k=5$    & $k=10$   & $k=20$ & \\\hline\hline
%		& & $HR@5$    & $HR@10$   & $HR@20$   & $NDCG@5$    & $NDCG@10$   & $NDCG@20$ & \\\hline\hline
		\multirow{10}{*}{\rotatebox[origin=c]{0}{Gowalla}} &Popularity& ItemPop & 0.2003 & 0.2785 & 0.3739 & 0.1099 & 0.1350 & 0.1591 &
		+263.2\%\\\cline{2-10}
		&\multirow{3}{*}{\rotatebox[origin=c]{0}{\specialcell{Linear Models}}}& MF-BPR & 0.6284 & 0.7480 & 0.8422 & 0.4825 & 0.5214 & 0.5454 & +9.5\%\\\cline{3-10}
		&& FISM & 0.6170 & 0.7381 & 0.8347 & 0.4742 & 0.5135 & 0.5381 & +11.1\%\\\cline{3-10}
		&& SVD++ & 0.6545 & 0.7634 & 0.8475 & 0.5111 & 0.5465 & 0.5679 & +5.6\%\\\cline{2-10}
		&\multirow{3}{*}{\rotatebox[origin=c]{0}{\specialcell{Neural Network\\Models}}}& MLP & 0.6359 & 0.7590 & 0.8535 & 0.4802 & 0.5202 & 0.5443 & +9.0\%\\\cline{3-10}
		&& JRL & 0.6685 & 0.7747 & 0.8561 & 0.5270 & 0.5615 & 0.5821 & +3.4\%\\\cline{3-10}
		&& NeuMF & 0.6744 & 0.7793 & 0.8602 & 0.5319 & 0.5660 & 0.5865 & +2.6\%\\\cline{2-10}
		&\multirow{3}{*}{\rotatebox[origin=c]{0}{\specialcell{Our Models}}}& ConvNCF-MF & 0.6914 & \textbf{0.7936$^*$} & \textbf{0.8695$^*$} & 0.5494 & 0.5826 & 0.6019 & -\\\cline{3-10}
		&& ConvNCF-FISM & 0.6716 & 0.7767 & 0.8572 & 0.5312 & 0.5654 & 0.5859 & -\\\cline{3-10}
		&& ConvNCF-SVD++ & \textbf{0.6949$^*$} & 0.7930 & 0.8657 & \textbf{0.5532$^*$} & \textbf{0.5851$^*$} & \textbf{0.6036$^*$} & -\\\hline\hline
		
		\multirow{10}{*}{\rotatebox[origin=c]{0}{Yelp}} &Popularity& ItemPop & 0.0710 & 0.1147 & 0.1732 & 0.0365 & 0.0505 & 0.0652 & +197.0\%\\\cline{2-10}
		&\multirow{3}{*}{\rotatebox[origin=c]{0}{\specialcell{Linear Models}}}& MF-BPR & 0.1752 & 0.2817 & 0.4203 & 0.1104 & 0.1447 & 0.1796 & +11.2\%\\\cline{3-10}
		&& FISM & 0.1852 & 0.2884 & 0.4262 & 0.1174 & 0.1505 & 0.1852 & +7.1\%\\\cline{3-10}
		&& SVD++ & 0.1893 & 0.2998 & 0.4360 & 0.1208 & 0.1562 & 0.1905 & +4.0\%\\\cline{2-10}
		&\multirow{3}{*}{\rotatebox[origin=c]{0}{\specialcell{Neural Network\\Models}}}& MLP & 0.1766 & 0.2831 & 0.4203 & 0.1103 & 0.1446 & 0.1792 & +11.1\%\\\cline{3-10}
		&& JRL & 0.1858 & 0.2922 & 0.4343 & 0.1177 & 0.1519 & 0.1877 & +6.0\%\\\cline{3-10}
		&& NeuMF & 0.1881 & 0.2958 & 0.4385 & 0.1189 & 0.1536 & 0.1895 & +4.9\%\\\cline{2-10}
		&\multirow{3}{*}{\rotatebox[origin=c]{0}{Our Models}}& ConvNCF-MF & 0.1978 & 0.3086 & 0.4430 & 0.1243 & 0.1600 & 0.1939 & -\\\cline{3-10}
		&& ConvNCF-FISM & 0.1925 & 0.3028 & 0.4423 & 0.1243 & 0.1598 & 0.1949 & -\\\cline{3-10}
		&& ConvNCF-SVD++ & \textbf{0.1991$^*$} & \textbf{0.3092$^*$} & \textbf{0.4457$^*$} & \textbf{0.1275$^*$} & \textbf{0.1629$^*$} & \textbf{0.1973$^*$} & -\\\hline
	\end{tabular}}
	
	\label{tab:performance}
\end{table*}

\section{Experiments}
\label{sec:expr}
To evaluate our proposal, we conduct experiments to answer the following research questions:
\begin{description}
	\item[RQ1] How do our ConvNCF perform compared with state-of-the-art recommendation methods?
	\item[RQ2] How do the embedding dimension correlations (captured by the outer product layer and convolution layers) contribute to the model performance? 
%	modeling embedding dimension correlations and using CNN layer helpful for learning from user-item interaction data and improving the recommendation performance?
%	\item[RQ3] How do the different choices of user embedding function  affect ConvNCF's performance?
	\item[RQ3] How do the key hyper-parameters (e.g., feature maps and pre-training) affect ConvNCF's performance?
\end{description}

\subsection{Experimental Settings}

\paragraph{Data Description}
We conduct experiments on two publicly accessible datasets: Yelp\footnote{https://github.com/hexiangnan/sigir16-eals} and Gowalla\footnote{http://dawenl.github.io/data/gowalla\_pro.zip}.

\textbf{Yelp}.~This is the Yelp Challenge data for user ratings on businesses.
We filter the dataset same as~\cite{he2016fast}. Moreover, we merge the repetitive ratings at different timestamps to the earliest one, so as to study the performance of recommending novel items to a user.
The final dataset obtains 25,815 users, 25,677 items, and 730,791 ratings.

\textbf{Gowalla}.~This is the check-in dataset from Gowalla, a location-based social network, constructed by~\cite{liang2016modeling} for item recommendation.
%Each interaction is a check-in record on a venue in Gowalla.
To ensure the quality of the dataset, we perform a modest filtering on the data, retaining users with at least two interactions and items with at least ten interactions.
The final dataset contains 54,156 users, 52,400 items, and 1,249,703 interactions. 

\paragraph{Evaluation Protocols}
For each user in the dataset, we holdout the latest one interaction as the testing positive sample, and then pair it with $999$ items that the user did not rate before as the negative samples.
Each method then generates predictions for these $1,000$ user-item pairs.
To evaluate the results, we adopt two metrics \textit{Hit Ratio} (HR) and \textit{Normalized Discounted Cumulative Gain} (NDCG), same as  \cite{he2017neural}.
HR@$k$ is a recall-based metric, measuring whether the testing item is in the top-$k$ position (1 for yes and 0 otherwise).
NDCG@$k$ assigns higher scores to the items within the top $k$ positions of the ranking list.
To eliminate the effect of random oscillation, we report the average scores of the last ten epochs upon convergence. 

\paragraph{Baselines}
To justify the effectiveness of our method, we compare with the following methods:

\begin{description}[style=unboxed,leftmargin=*]
\item[ItemPop] ranks the items based on their popularity, which is calculated by the number of interactions. It is always taken as a benchmark for recommender algorithms.

\item[MF-BPR~\cite{rendle2009bpr}] optimizes the matrix factorization model with the pairwise BPR ranking loss. It is a competitive user-based CF method.  

\item[FISM~\cite{kabbur2013fism}] replaces the user ID embedding with the item-based embedding function of Equation~\ref{eq:fism}. 
It is a competitive item-based CF method.

\item[SVD++~\cite{koren2008factorization}] combines the user embedding design of MF and FISM, as formulated in \ref{eq:svdpp}. It is a strong CF model that scores the best single model in the Netflix challenge. 

%It is the state-of-the-art hybrid model. 

%It upgrade the target generator, one MF-BPR model, via an adversarial loss provided by the discriminator, another MF-BPR. 
\item[MLP~\cite{he2017neural}] is a NCF method that feeds the concatenation of user embedding and item embedding into the standard MLP for learning the interaction function. As no interaction between user embedding and item embedding is explicitly modeled, this model can be inferior to the MF model~\cite{he2017neural}. 

\item[JRL~\cite{zhang2017joint}] is a NCF method that places a MLP above the element-wise product of user embedding and item embedding. It enhances GMF~\cite{he2017neural} by placing multiple hidden layers above the element-wise product, while GMF directly outputs the prediction score from the element-wise product. 

\item[NeuMF~\cite{he2017neural}] is a state-of-the-art method for item recommendation, which ensembles GMF and MLP to learn the user-item interaction function.
\end{description}

The architecture of ConvNCF is shown in Figure~\ref{fig:newframe}. Above the outer product layer, there are six convolutional layers. Each of them has $32$ convolutional kernels with stride 2 and no padding.
We implement the three ConvNCF methods as introduced in Section~\ref{sec:methods}, which have the same architecture setting with difference only in the user embedding function. 
%and mark them with their embedding types as \textbf{ConvNCF-MF}, \textbf{ConvNCF-FISM} and \textbf{ConvNCF-SVD++}.

\paragraph{Parameter Settings}
Our methods are implemented in Tensorflow\footnote{Available at: \href{https://github.com/duxy-me/ConvNCF}{https://github.com/duxy-me/ConvNCF}.}.  We randomly holdout 1 training interaction for each user as the validation set to tune hyper-parameters.
For a fair comparison, all models apply the same setting in terms of the model size and optimization: the embedding size is set to 64 and all models are optimized with the BPR loss using mini-batch Adagrad. 
%We evaluate the three instantiations of ConvNCF as illustrated in Figure~\ref{fig:embs}.
%SPACE
%, which has 6 convolutional layers and each layer has the same number of feature maps (the default setting is 32). 
%The outputs of each layer have half width and height of its previous layer, and the last convolutional layer outputs a vector indicating the abstractive user-item interaction feature.
%The regularization coefficients are separately tuned for the embedding layer, convolutional layers, and output layer in the range of $[10^{-3}, 10^{-2}, ..., 10^2]$.  
%we set the embedding size as 64 for all models and optimize them with the same BPR loss using .
For MLP, JRL, and NeuMF that have multiple fully connected layers, we tuned the number of layers from 1 to 3 following the tower structure of \cite{he2017neural}. 
For all models besides MF, we pre-train their embedding layers using MF embeddings, and the $L_2$ regularization for each method has been fairly tuned. 
%We pretrain all models with MF-BPR embeddings for around $1,500$ iterations.

Due to the strong representation ability of ConvNCF, it is prone to overfitting. As such, tuning the regularization has a large impact on its performance. We divide the parameters of ConvNCF into two groups: embedding parameters ($\Theta_U$ and $\Theta_I$) and CNN parameters ($\Theta_{CNN}$ and $\textbf{w}$), and tune the regularization coefficient and learning rate separately for the two groups. For the embedding parameters, the optimal learning rate (under Adagrad) is around $0.001$ and $0.005$, and the regularization coefficients (i.e., $\lambda_1$, $\lambda_2$) are searched in $[0,10^{-7},10^{-6}]$. 
For the CNN parameters, the optimal learning rate is around 0.01 and 0.05, and the regularization coefficients (i.e., $\lambda_3$, $\lambda_4$) are searched in $[10^{-2},10^{-1},1,10]$. 

%In Equation~\ref{eq:bprconv} we defined $\lambda_1$, $\lambda_2$ and $\lambda_3$ for regularizing the parameters. Since $\Theta_u$ and $\Theta_i$ are mostly equivalent in linear and our model, to simplify the training process, we let $\lambda_1 \equiv \lambda_2$. Similarly, we let $\lambda_3 \equiv \lambda_4$. Moreover, assuming that the pretrained part would have minor changes while the other part would learn much, we set different learning rates and regularizations for them. Usually, we set a small learning rate ranged in $[0.001,0.005]$ for $\Theta_u$ and $\Theta_i$, and set a larger learning rate ranged in $[0.01,0.05]$ for $\Theta_{CNN}$ and $\textbf{w}$. The coefficient $\lambda_1$ and  $\lambda_2$ are ranged in $[0,10^{-7},10^{-6}]$ and $\lambda_3$ and  $\lambda_4$ are ranged in $[10^{-2},10^{-1},1,10]$. Considering the fully-connected layer would have unexpected over-fitting, the regularization on the parameters in last layer could be tuned independently.

\subsection{Performance Comparison (RQ1)}
\begin{figure}[t!]
		\centering
		\includegraphics[width=\linewidth]{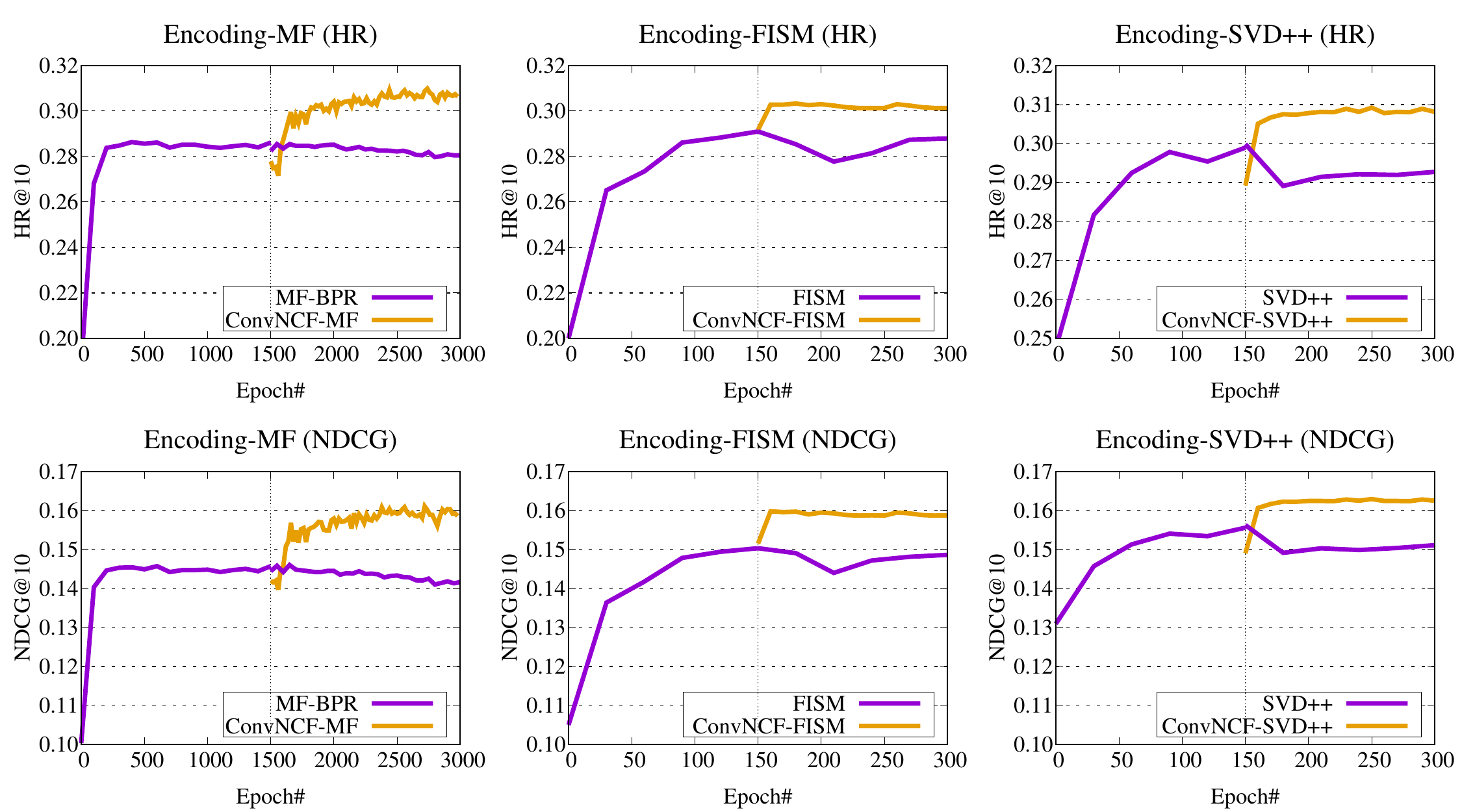}
		\caption{HR@10 and NDCG@10 of ConvNCF models and corresponding embedding models~(\ie MF-BPR~\cite{rendle2009bpr}, SVD++~\cite{koren2008factorization} and FISM~\cite{kabbur2013fism}) on Yelp. The results verify that ConvNCF could effectively discover the inherent correlations between embedding dimensions.}
		\label{fig:continue}
	\end{figure}

	Table~\ref{tab:performance} shows the Top-$k$ recommendation performance on both datasets where $k$ is set to 5, 10, and 20.
	We have the following key observations:
	\begin{itemize}[leftmargin=*]
	    \item Our ConvNCF methods achieve the best performance in general, and obtain high improvements over the compared methods. Particularly, each ConvNCF instantiation outperforms its linear counterpart in all settings (e.g., ConvNCF-MF vs. MF, ConvNCF-FISM vs. FISM). 
	    This justifies the utility of our ConvNCF design --- using outer product (and CNN) to capture second-order (and high-order) correlations among embedding dimensions. 
	    %\item With the same definition of $f^U(u)$ and $f^I(i)$, ConvNCF always outperform the corresponding linear models. That verifies it is surely necessary to model the correlations among embedding dimensions.
		\item Among the three ConvNCF methods, ConvNCF-SVD++ is the strongest and achieves the best performance in most settings. This demonstrates the utility of designing better embedding function for ConvNCF. Moreover, we find that ConvNCF-FISM achieves weak performance on Gowalla. We hypothesize the reason is that the item-based user representation mechanism does not work well for the Gowalla dataset, since FISM also underperforms MF substantially on Gowalla. Another obsrevation is that initialized with the pretrained embeddings, ConvNCF would significantly improve the performances in few epochs~(Figure~\ref{fig:continue}). In such few epochs, the embeddings would not change much, that implies that modeling the embedding correlations plays a significant role for the state-of-the-art performances.  
		%The three instantiations of ConvNCF obtain well performances on both datasets, but the performance of ConvNCF-FISM on Gowalla is a bit worse. Noticing that the FISM performs worse than MF on Gowalla, we believe that the item-based feature does not fit the dataset Gowalla well. Besides, ConvNCF-SVD++ obtains most of the best results under various indicators. That indicates the high quality of SVD++ embedding.
 	    %By operating outer product over the embeddings of users and items, ConvNCF is capable of uncovering the second-order interactions among feature dimensions. Subsequently, performing CNN layers can learn higher-order correlations among the embedding dimensions from locally to globally. It justifies the effectiveness of our proposed method.
		\item JRL consistently outperforms MLP by a large margin on both datasets.
		This indicates that, explicitly modeling the correlations of embedding dimensions is rather helpful for the followup learning of the hidden layers, even modeling the simple correlations that assume the dimensions are independent. 
		Meanwhile, it also reveals the practical difficulties to train MLP well, although it has strong representation ability in principle~\cite{hornik1991approximation}.
		\item The relatively weak performances of MF and FISM reflect that simple inner product of the user and item embeddings is far insufficient to depict the complex patterns within the user-item interactions. Moreover, MF outperforms FISM on Gowalla but underperforms FISM on Yelp. This implies that which embedding function works better is dependent on the properties of the dataset.
		%It again justifies the inner product easily leads to information loss, compared to the element-wise product and our proposed outer product.
		Lastly, ItemPop achieves the worst performance, verifying the importance of considering users' personalized preferences in the recommendation.
	\end{itemize}
	
	\subsection{Modeling Embedding Dimension Correlations (RQ2)}
	%In the section, we purpose to justify how the proposed outer product and CNN affect the recommendation performance.
	
	%Table~\ref{tab:performance} demonstrates that with the same definition of $f^U(u)$ and $f^I(i)$,, ConvNCF always obtains better performances with large margin. That verifies that ConvNCF is effective to sufficiently dig the potential information in embeddings. 
	Through overall performance comparison, we have shown the strength of ConvNCF. Next, we conduct more experiments to verify the utility of modeling embedding dimension correlations, more concretely, the efficacy of outer product and CNN in ConvNCF. We choose ConvNCF-MF as an example to demonstrate the usefulness of outer product (for brevity, we term it as ConvNCF in this subsection only). 

	%Subsequently, we conduct two experiments and record the performances to further explore the efficacy of the two parts, outer product and CNN.
	
	\subsubsection{Efficacy of Outer Product}
\begin{figure}[t!]
		\centering
		\includegraphics[width=0.8\linewidth]{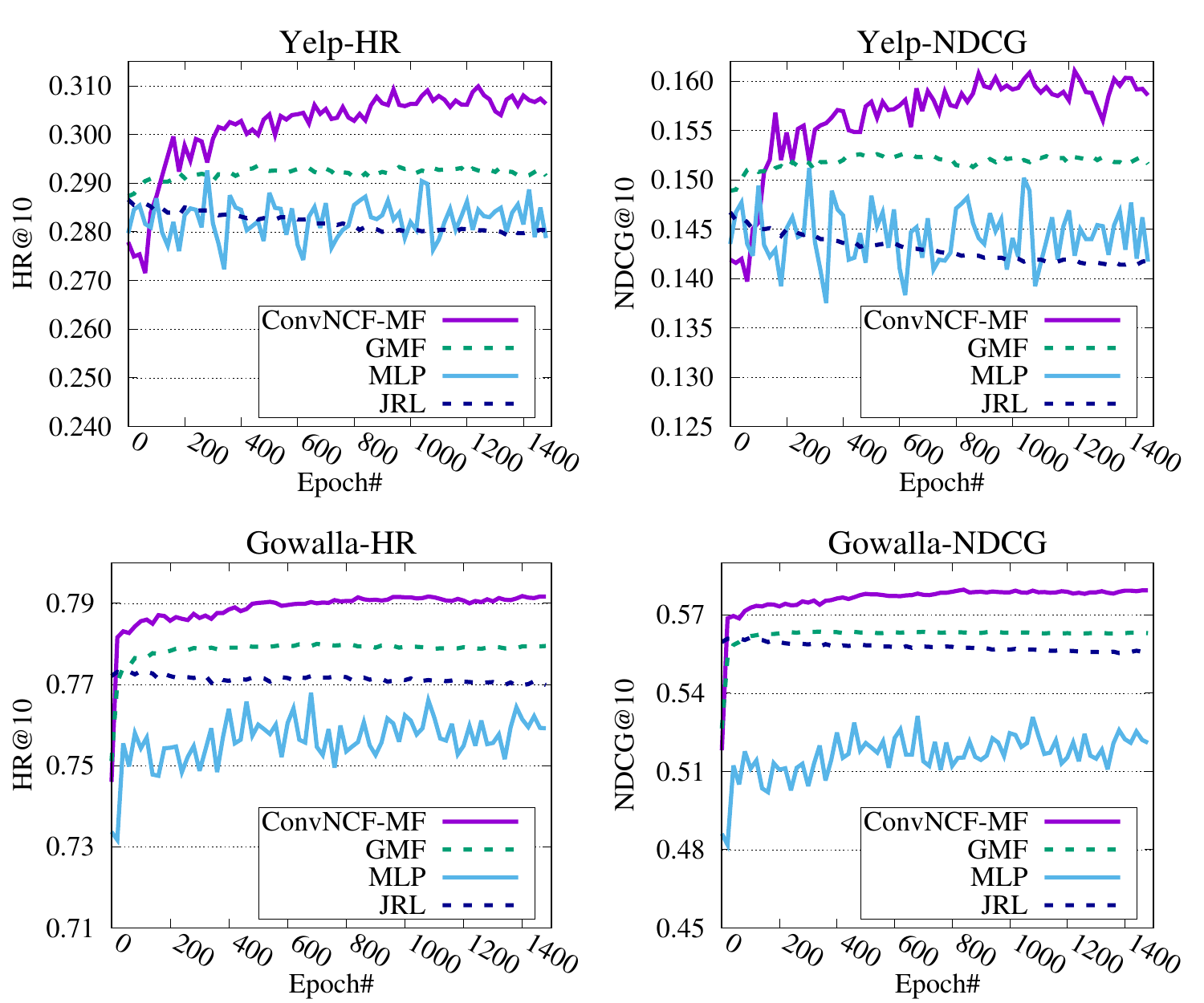}
		\caption{HR@10 and NDCG@10 of applying different operations above the embedding layer in each epoch (GMF and JRL use element-wise product, MLP uses concatenation, and ConvNCF uses outer product). }
		\label{fig:ovi}
	\end{figure}

	Besides outer product, another two choices that are commonly used in previous work are concatenation and element-wise product.
	\dxy{It is worth mentioning that element-wise product, concatenation, and outer product operations have essentially different structures, since the outer product operation is performed based on a matrix while the other two operations are performed based on vectors. To conduct a fair evaluation, we implement the state-of-the-art models representing element-wise product and the concatenation style learning methods according to ~\cite{zhang2017joint,he2017neural}.}
	As such, we compare the training progress of ConvNCF with GMF and JRL (which use element-wise product), and MLP (which uses concatenation). 
	As can be seen from Figure \ref{fig:ovi}, ConvNCF outperforms other methods by a large margin on both datasets, verifying the positive effect of using outer product above the embedding layer. 
	Specifically, the improvements over GMF and JRL demonstrate that explicitly modeling the correlations between different embedding dimensions are useful.
	Lastly, the rather weak and unstable performance of MLP imply the difficulties to train MLP well, especially when the low-level has fewer semantics about the feature interactions. This is consistent with the recent finding of \cite{NFM} in using MLP for sparse data prediction. %\vspace{+5pt}. 
	%Jointly analyzing the performance in Figure~\ref{fig:ovi}, we have the following findings:
	%\begin{itemize}[leftmargin=*]
 %Clearly, ConvNCF outperforms the other methods by a large margin w.r.t. HR@10 and NDCG@10 on different datasets.
		%While MLP may utilize the stacked hidden layers to model the feature interactions in an implicit way, and GMF and JRL model the partial dimension correlations via element-wise product, neither of them explicitly presents all the dimension correlation. This makes them only consider partial information and less their expressiveness. It again verifies the effectiveness and expressiveness of the outer product.
		%\item Owing to more user-item interactions existing in Yelp, many baselines oscillate intensely, specifically for GMF.
		%Such data may contribute to its unsatisfactory performance due to its limited expressiveness.
		%In contrast, ConvNCF is more stable and oscillates a bit, verifying that the outer product can effectively reduce information loss and prevent the model from overfitting.
	%\end{itemize}
	
	\subsubsection{Efficacy of CNN} 
\begin{figure}[t!]
		\centering
		\includegraphics[width=0.8\linewidth]{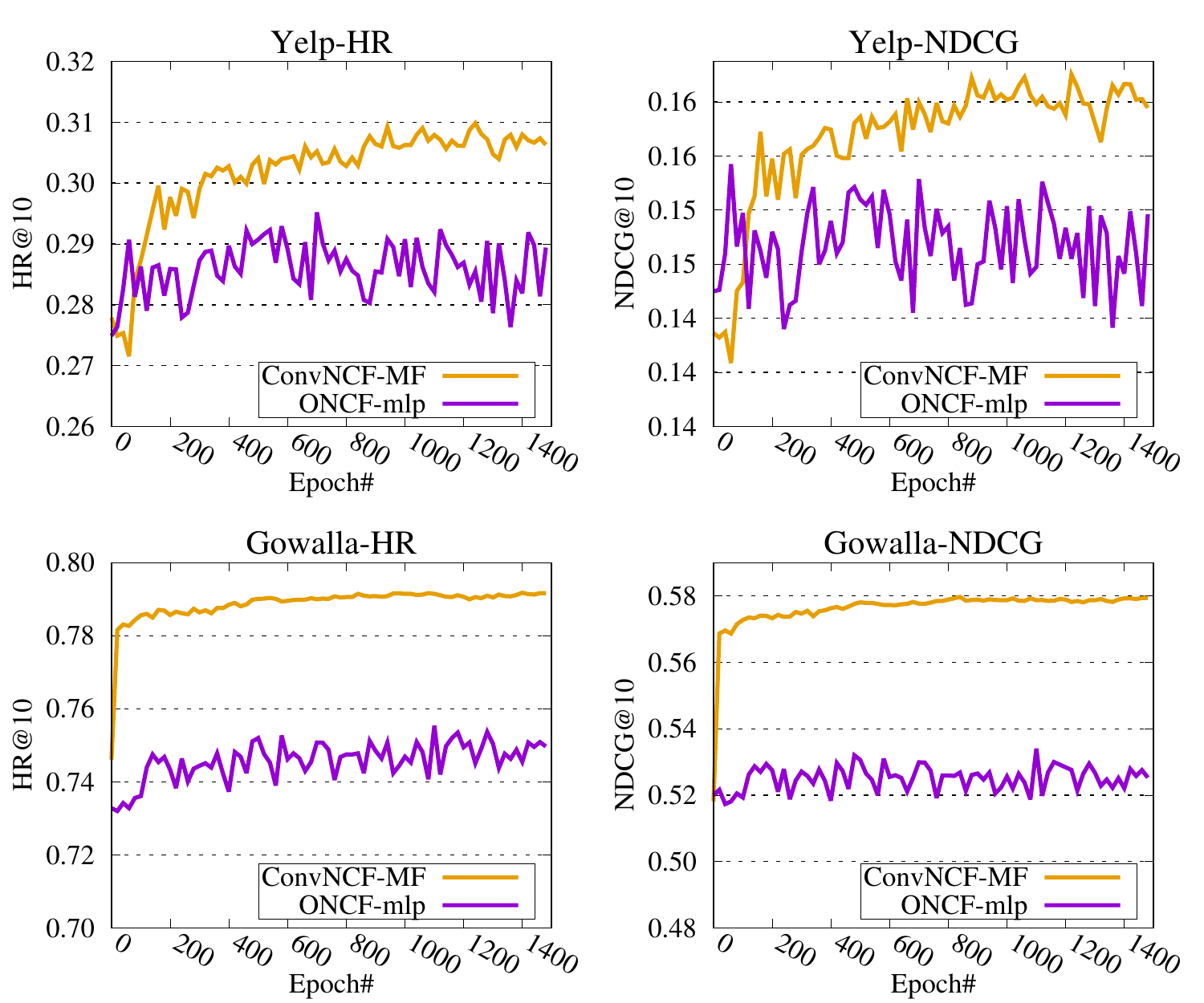}
		\caption{HR@10 and NDCG@10 of using different hidden layers above the interaction map (ConvNCF uses a 6-layer CNN and ONCF-mlp uses a 3-layer MLP). }
		\label{fig:cnn}
	\end{figure}

\
	In order to verify the effectiveness of CNN over MLP, we make a fair comparison by training them based on the same 2D interaction map. 
	Specifically, we first flatten the interaction map as a $K^2$ dimensional vector, and then place a 3-layer MLP above it. We term this method as ONCF-mlp.  Figure~\ref{fig:cnn} compares its performance with ConvNCF in each epoch. We can see that ONCF-mlp performs much worse than ConvNCF, in spite of the fact that it uses much more parameters (1000 times more) than ConvNCF. Another drawback of using such many parameters in ONCF-mlp is that it makes the model rather unstable, which is evidenced by its large variance in epoch. 
	In contrast, our ConvNCF achieves much better and stable performance by using the locally connected CNN. These empirical evidence provide support for our motivation of designing ConvNCF and our discussion of MLP's drawbacks in Section~\ref{subsubsec:conv_mlp}.
	
	%To validate the influence of CNN, we use MLP as a substitute and conduct a comparison experiment. Particularly, we conduct the MLP model over the feature mappings, (i.e., the outputs of the outer product), denoted as ONCF-mlp. Figure~\ref{fig:cnn} exhibits the performance comparison w.r.t. HR@10 and NDCG@10 over two datasets. We have the following insights:
	%\begin{itemize}[leftmargin=*]
		%\item Clearly, replacing the CNN part with MLP hurts the expressiveness adversely and degrades the recommendation performance. ONCF-mlp only relies on the hidden layers to automatically learn the feature interactions, which fails to explicitly model the dimension correlation.
		%Taking advantage of CNN, ConvNCF can fetch more abstract patterns when going deeper and capture the spatial dimension correlation from locally to globally.
	%	\item The learning process of ConvNCF is much more stable than that of ONCF-mlp, justifying the robustness of our proposed methods.
	%\end{itemize}
	
	\subsection{Hyperparameter Study (RQ3)}
	%We empirically analyze the influences of several hyperparameters, such as the number of convolutional filters, regularization, and dropout ratio, in the following section.
	
	\subsubsection{Impact of Feature Map Number} 
\begin{figure}[t!]
	\centering
	%\subfigure{\label{fig:layer_1}\includegraphics[width=0.47\linewidth]{./images/layer/hr.pdf}}
	%\subfigure{\label{fig:layer_2}\includegraphics[width=0.47\linewidth]{./images/layer/ndcg.pdf}}
	\includegraphics[width=0.8\linewidth]{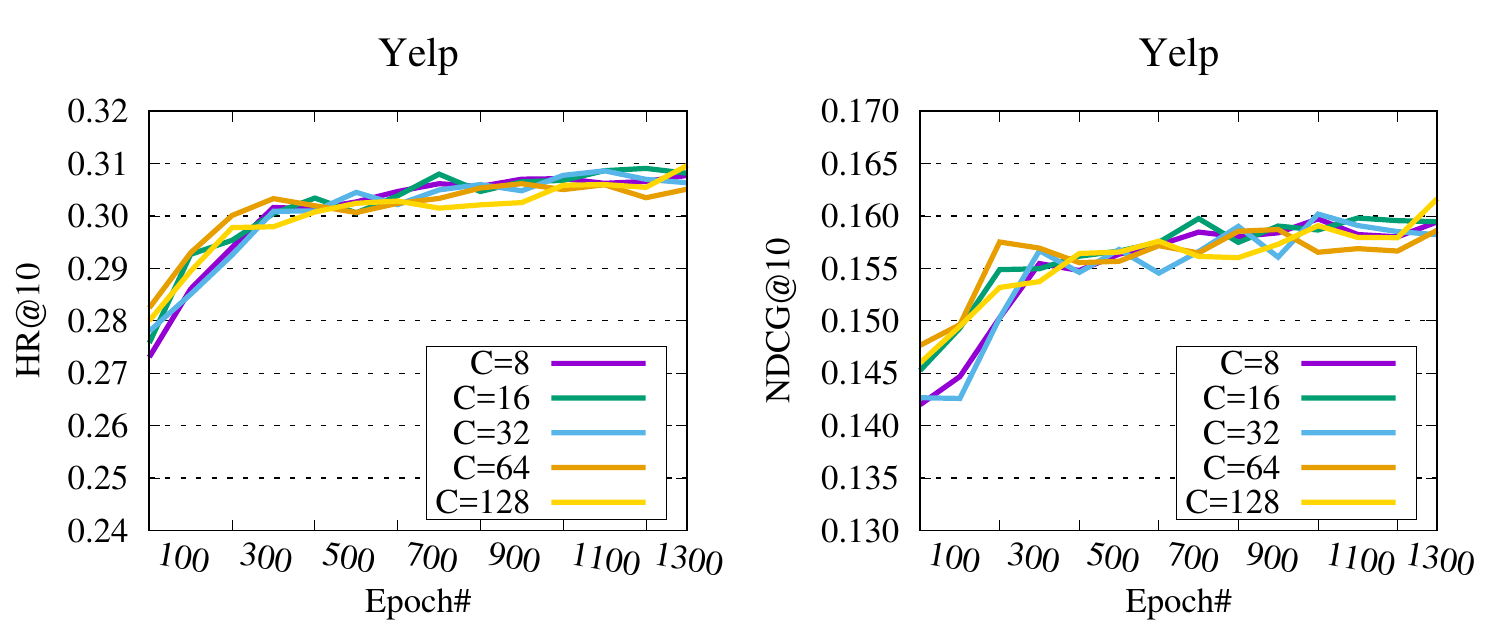}
	\caption{Performance of ConvNCF-MF w.r.t. different numbers of feature maps per convolutional layer (denoted by $C$) in each epoch on Yelp. }
	\label{fig:layer}
\end{figure}

The number of feature maps in each CNN layer affects the representation ability of our ConvNCF. %Theoretically speaking, using more feature maps makes ConvNCF more expressive. 
	%Here we empirically investigate how varying the number of feature maps impacts ConvNCF's performance. 
	Figure \ref{fig:layer} shows the performance of ConvNCF-MF with respect to different numbers of feature maps.
	%(note that here we use the same number for all convolutional layers). 
	We can see that all the curves increase steadily and finally achieve similar performance, though there are some slight differences on the convergence curve. This reflects the strong expressiveness and generalization of using CNN above the interaction map since dramatically increasing the number of parameters of a neural network does not lead to overfitting. Consequently, our model is very suitable for practical use.

%	To verify this, we further show ConvNCF's performance with different dropout ratios in Table~\ref{tab:dropout}. We can see that even no dropout used, it performs quite well and tuning the dropout ratio only leads to slight improvements. As such, we conclude that  our proposed ConvNCF has remarkable generalization ability and robustness, and thus is very suitable for practical use.

%\begin{table}
%	\centering
%	\caption{Performance of ConvNCF w.r.t. dropout ratio $\rho$.}\vspace{-10pt}
%	\label{tab:dropout}
%	\resizebox{0.42\textwidth}{!}{%
%		\begin{tabular}{|c|c|c||c|c|}
%			\hline
%			& \multicolumn{2}{c||}{\textbf{Gowalla}} & \multicolumn{2}{c|}{\textbf{Yelp}} \\\hline
%			$\mathbf{\rho}$ & \textbf{HR}@10 & \textbf{NDCG}@10 & \textbf{HR}@10 & \textbf{NDCG}@10\\\hline\hline
%			0.4 & 0.7924&0.5821 & 0.3005 & 0.1564 \\\hline
%			0.2 & 0.7940 &0.5826 & 0.3008 & 0.1565\\\hline
%			0.1 & 0.7938& 0.5827& 0.3065 & 0.1586 \\\hline
%			0 & 0.7915 & 0.5797 &0.3086& 0.1600 \\\hline
%		\end{tabular}}\vspace{-10px}
%\end{table}

\subsubsection{Pretraining on Embeddings.} 
\begin{figure}[t!]
	\centering
	\includegraphics[width=\linewidth]{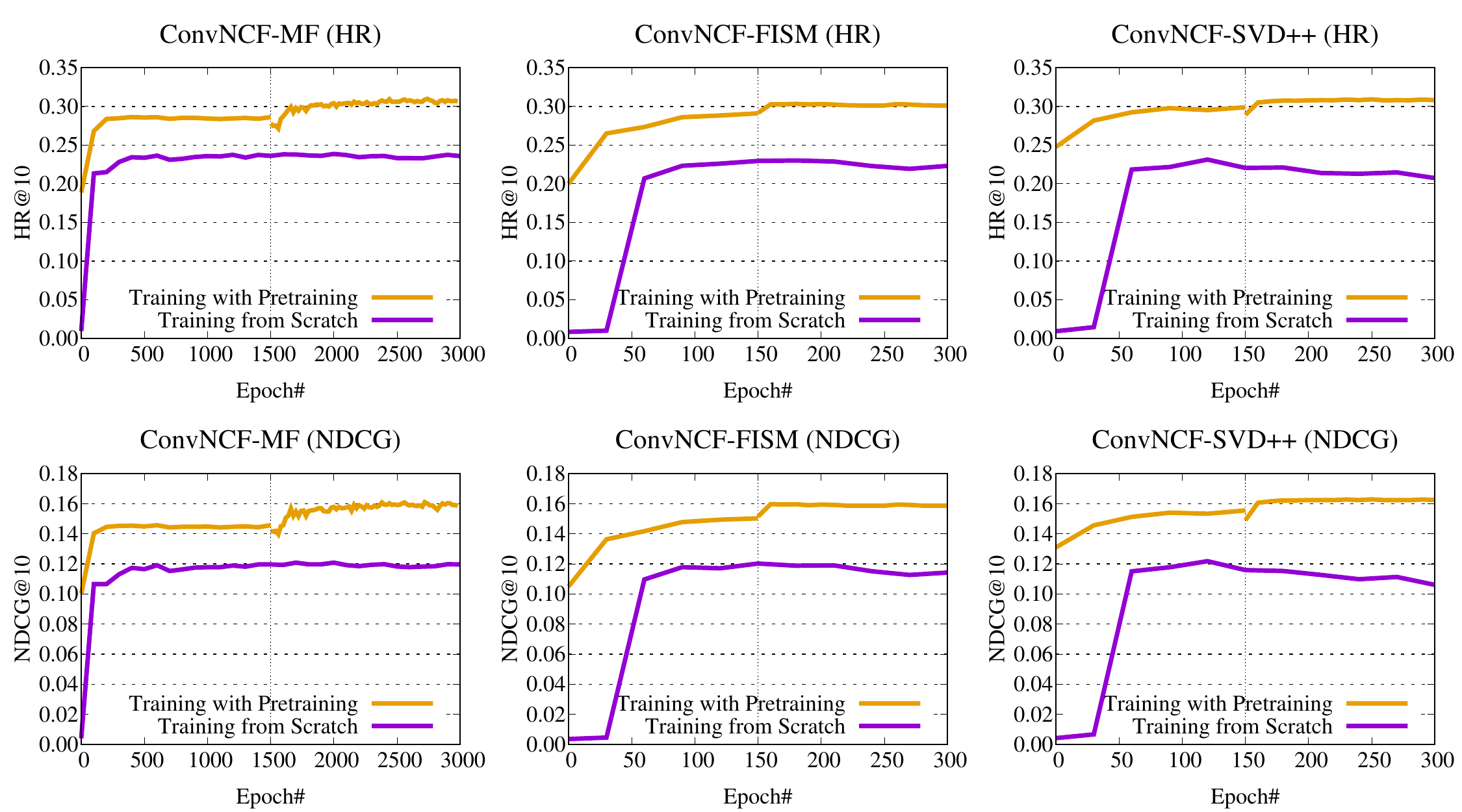}
	\caption{HR@10 and NDCG@10 on Yelp via different training tricks: training from scratch and training with pretraining. The results demonstrate that pretraining the embeddings with original models is helpful to efficiently obtain the well-performed ConvNCF.}
	\label{fig:pretrain}
\end{figure}

In our experiments, all the embeddings used in ConvNCF are initialized by the pretrained parameters. We here compare the effect of training with pretraining and training from scratch. As shown in Figure~\ref{fig:pretrain}, there are two curves. The orange one indicates the performance training with pretraining, and the purple one indicates the performance training from scratch. The orange curve at the left side of the dashed line presents the status pretraining the models and the rest presents the status training ConvNCF. Due to the simplicity of original models, the pretraining processs increase the performances soon. Based on these embeddings, ConvNCF obtains the state-of-the-art performance, which re-emphasizes the importance of modeling the dimension correlations. In contrast, training from scratch also grasps some information but due to more complex structure, it requires more efforts to achieve the nice performance.

\section{Conclusion}
In this paper, we presented a new neural network framework ConvNCF for collaborative filtering by modeling  the embedding dimension correlations. Moreover, we  proposed three variants of ConvNCF.
% We modeled the embedding dimension correlations, presented a new neural network framework for collaborative filtering, named ConvNCF, and implemented three instantiations of ConvNCF. 
ConvNCF is able to capture latent relations between embedding dimensions via the outer product operation. 
% which results in a semantic-rich interaction map. 
To learn more accurate preference, we used multiple convolution layers on top of the interaction map. Extensive experiments on two real-world datasets demonstrated that modeling the embedding dimension correlations was helpful for CF tasks. We further showed that ConvNCF outperforms state-of-the-art methods in the top-$k$ item recommendation task. 

% and ConvNCF uses the embedding more adequately to outperform state-of-the-art methods in top-$k$ recommendation. 

In future, we will explore more advanced CNN techniques such as attentive mechanism~\cite{lin2017structured} and residual learning~\cite{he2016deep} to learn higher-level representations.
Moreover, we will extend ConvNCF to content-based recommendation scenarios~\cite{he2016vbpr,Yu:2018:ACR} since the item features have richer semantics than just an ID. Particularly, we are interested in building recommender systems for multimedia items like images and videos, and textual items like news.

%
% The acknowledgments section is defined using the "acks" environment (and NOT an unnumbered section). This ensures
% the proper identification of the section in the article metadata, and the consistent spelling of the heading.
\begin{acks}
This work was partially supported by the National Key Research and Development Program of China under Grant 2016YFB1001001, the National Natural Science Foundation of China (Grant No. 61772275, 61732007, 61720106004 and U1611461), the Natural Science Foundation of Jiangsu Province (BK20170033), and the National Science Foundation of China - Guangdong Joint Foundation (Grant No. U1401257). This research is also part of NExT++, supported by the National Research Foundation, Prime Ministers Office, Singapore under its IRC@Singapore Funding Initiative.
\end{acks}

%
% The next two lines define the bibliography style to be used, and the bibliography file.
\bibliographystyle{ACM-Reference-Format}
\bibliography{main}

\end{document}